\documentclass[aps,prb,twocolumn,superscriptaddress,showpacs,floatfix,longbibliography]{revtex4-1}
\usepackage{amsmath,amssymb}

\usepackage[utf8]{inputenc}
\usepackage[T1]{fontenc}
\usepackage{xcolor}
\usepackage{booktabs}
\usepackage{array}
\usepackage{diagbox}
\IfFileExists{newtxtext.sty}
   {\usepackage{newtxtext,newtxmath}}
   {\IfFileExists{stix.sty}
      {\usepackage{stix}}
      {\IfFileExists{mathptmx.sty}
      {\usepackage{mathptmx}}{} } }

\usepackage{textcomp}

\usepackage{bm}

\IfFileExists{siunitx.sty}{\usepackage{booktabs,siunitx}}{}

\pdfoutput=1
\usepackage{color}
\definecolor{LinkColor}{rgb}{0.256,0.439,0.588}
\usepackage{hyperref}
\hypersetup{
   pdfauthor={abc},
   pdftitle={paper},
   colorlinks=true,
   citecolor=blue,
   linkcolor=blue,
   urlcolor=blue
}

\renewcommand{\vec}[1]{\mathbf{#1}}

\usepackage{pifont}

\usepackage[pdftex]{graphicx}
\DeclareGraphicsExtensions{.pdf,.jpeg,.png,.eps}
\usepackage{epstopdf}
\usepackage{soul}

\begin{document}
	\bibliographystyle{apsrev4-1}
	\title{Bose metal in exactly solvable model with infinite-range Hatsugai-Kohmoto interaction}
	
	\author{Wei-Wei Yang}
	\affiliation{Key Laboratory of Quantum Theory and Applications of MoE $\&$ School of Physical Science and Technology, Lanzhou University, Lanzhou 730000, People Republic of China} %
	\author{Hong-Gang Luo}
	\affiliation{Key Laboratory of Quantum Theory and Applications of MoE $\&$ School of Physical Science and Technology, Lanzhou University, Lanzhou 730000, People Republic of China} %
	\affiliation{Beijing Computational Science Research Center, Beijing 100084, China}%
    \affiliation{Lanzhou Center for Theoretical Physics, Key Laboratory of Theoretical Physics of Gansu Province, Lanzhou University, Lanzhou 730000, People Republic of China}

	\author{Yin Zhong}
    \email{zhongy@lzu.edu.cn}
\affiliation{Key Laboratory of Quantum Theory and Applications of MoE $\&$ School of Physical Science and Technology, Lanzhou University, Lanzhou 730000, People Republic of China}
    \affiliation{Lanzhou Center for Theoretical Physics, Key Laboratory of Theoretical Physics of Gansu Province, Lanzhou University, Lanzhou 730000, People Republic of China}
	
\begin{abstract}
	In a conventional boson system, the ground state can either be an insulator or a superfluid (SF) due to the duality between particle number and phase.
	This paper reveals that the long-sought Bose metal (BM) state can be realized in an exactly solvable interacting bosonic model, i.e. the Bose-Hatsugai-Kohmoto (BHK) model, which acts as the nontrivial extension of Bose-Hubbard (BH) model.
By tuning the parameters such as bandwidth $W$, chemical potential $\mu$, and interaction strength $U$, a BM state without any symmetry-breaking can be accessed for a generic $W/U$ ratio, while a Mott insulator (MI) with integer boson density is observed at small $W/U$.
 The quantum phase transition between the MI and BM states belongs to the universality class of the Lifshitz transition, which is further confirmed by analyzing the momentum-distribution function, the Drude weight, and the SF weight.
Additionally, our investigation at finite temperature reveals similarities between the BM state and the Fermi liquid, such as a linear-$T$ dependent heat capacity ($Cv\sim \gamma T$) and a saturated charge susceptibility ($\chi_{c} \sim $ constant) as $T$ approaches zero.
Comparing the BM state with the SF state in the standard BH model, we find that the key feature of the BM state is a compressible total wavefunction accompanied by an incompressible zero-momentum component.
Given that the BM state prevails over the SF state at any finite $U$ in the BHK model, our work suggests the possibility of realizing the BM state with on-site repulsion interactions in momentum space.
\end{abstract}
	
	\date{\today}
	
	\maketitle
	
	%\tableofcontents
	
\section{\label{sec1:level1}Introduction}	

In traditional interacting boson systems, bosons manifest as eigenstates of either the phase operator or the particle number operator, which correspond to superfluid (SF) \cite{fetter_quantum_1976} and insulating states \cite{greiner_quantum_2002}, respectively.
Within the well-known Bose-Hubbard (BH) model incorporating on-site interaction, bosons with weak interaction typically yield a SF state, while stronger interactions coupled with integer boson filling result in a bosonic Mott insulator (MI) that reinstates $U(1)$ symmetry \cite{Sachdev}.
Introducing disorder can trigger the appearance of a Bose glass exhibiting replica symmetry breaking \cite{fisher_boson_1989}.
However, in both circumstances, a metallic state does not readily emerge during the SF-MI transition.
Furthermore, the No-go theorem—specifically the 'Gang of four' scaling theory of localization—states the absence of metallic states in a two-dimensional system that involves disorder \cite{abrahams_scaling_1979}.
Consequently, identifying the presence of a metallic state in low-dimensional boson systems appears elusive and challenging.

Remarkably, an anomalous metallic state exhibiting residual resistance significantly lower than the quantum resistance ($h/e^{2}$) at low temperatures has been observed experimentally \cite{PhysRevB.40.182,PhysRevLett.76.1529,PhysRevLett.82.5341,PhysRevLett.88.037004,Eley2012,PhysRevLett.111.067003,PhysRevB.87.024509,Han2014,doi:10.1126/science.1259440,doi:10.1126/sciadv.1700612,doi:10.1126/science.1259440,RevModPhys.91.011002,Bottcher2018,Chen2018,doi:10.1126/science.aax5798,Tsen2016}.
This metallic state is speculated to occur between superconducting and insulating states under specific conditions: the tuning of thinness, magnetic field, or gate voltage in superconducting films, Josephson-junction arrays, and superconducting islands \cite{RevModPhys.91.011002,PhysRevLett.80.3352,PhysRevB.63.125322,PhysRevLett.95.077002,PhysRevLett.67.3302,PhysRevB.60.1261,doi:10.1126/science.1088253,PhysRevB.64.132502,PhysRevB.77.214523,PhysRevB.93.205116,FEIGELMAN1998107,PhysRevLett.89.027001,PhysRevB.73.214507,PhysRevLett.95.077002,PhysRevB.94.054502,PhysRevB.93.205116,PhysRevB.77.214523, PhysRevB.64.134511,PhysRevB.54.10081,chen2021universal}.
Further experiments have observed the charge-$2e$ quantum oscillation \cite{doi:10.1126/science.aax5798} and vanished Hall resistivity \cite{doi:10.1126/sciadv.1700612}, which suggest that bosonic particles, i.e. the Cooper pair formed by two electrons, should play a decisive role in the anomalous metallic state between superconducting and insulating states. Therefore, the metallic states with bosonic nature has been identified as the Bose metal (BM) \cite{RevModPhys.91.011002,doi:10.1126/science.1088253}, which could potentially provide an explanation for these unusual metallic states themselves.

BM, as a common trend in the field of condensed matter physics, has stimulated numerous interesting theories, such as the phase glass, fractionalization, dissipation effect, vortex liquid, quantum Boltzmann theory, and the composite fermions \cite{PhysRevB.48.16641,PhysRevB.75.235116,PhysRevB.64.052507,PhysRevLett.82.5341,RevModPhys.91.011002,PhysRevLett.80.3352,PhysRevB.63.125322,PhysRevLett.95.077002,PhysRevLett.67.3302,PhysRevB.60.1261,doi:10.1126/science.1088253,PhysRevB.64.132502,PhysRevB.77.214523,PhysRevB.93.205116,FEIGELMAN1998107,PhysRevLett.89.027001,PhysRevB.73.214507,PhysRevLett.95.077002,PhysRevB.94.054502,PhysRevB.93.205116,PhysRevB.77.214523, PhysRevB.64.134511}.  An intriguing proposal suggests BM may exhibit behaviors akin to a Fermi liquid, with a Bose surface (comparable to the Fermi surface for fermions) where the excitation energy diminishes and gapless excitations naturally occur\cite{PhysRevB.48.16641,PhysRevB.75.235116}. Contrary to the Fermi surface, the Bose surface does not demarcate the boundary between occupied and unoccupied states.
However, despite many years of research, due to the complex interplay of correlation, disorder and magnetic field, the existing theories on this subject are often based on approximations that are difficult to control, such as mean-field decoupling, Gaussian effective action and slave-particle splitting with only large-$N$ limit. (Note however some numerical evidences on BM with Bose surface supplemented with frustrated interaction \cite{PhysRevB.78.054520,PhysRevLett.106.046402}.) Therefore, the underlying mechanisms behind the anomalous metal phenomenon are still not fully understood, which makes it an ongoing topic of investigation in condensed matter physics \cite{RevModPhys.91.011002}.

Given the challenges posed by BM states, we propose a simpler question: is it feasible to identify an exactly solvable model that demonstrates BM as its ground state?
Obviously, such thinking is motivated by recent progress on many solvable models, ranging from Kitaev's toric code, honeycomb lattice model to Sachdev-Ye-Kitaev model and Hatsugai-Kohmoto (HK) model \cite{doi:10.1143/JPSJ.61.2056,doi:10.1142/S0217984991000782,PhysRevLett.70.3339,PhysRevD.94.106002,RevModPhys.94.035004,RevModPhys.89.025003,PhysRevB.96.205104,PhysRevB.88.045109,PhysRevLett.118.266601,PhysRevLett.120.046401,KITAEV20032,KITAEV20062}.
These models have yielded intriguing quantum spin liquids and non-Fermi liquids, enhancing our understanding of spin liquids with Majorana fermion excitation and non-Fermi liquids without quasiparticle in the presence of dominant disordered interaction.

In the present study, we uncover a BM state within an exactly solvable model with infinite-range interaction, namely, the Bose-Hatsugai-Kohmoto (BHK) model \cite{CONTINENTINO1994619}.
It is crucial to note that the BHK model serves as the bosonic counterpart to the extensively researched HK model \cite{PhysRevB.54.5358,PhysRevB.97.195102,PhysRevD.99.094030,PhysRevB.103.024514,PhysRevB.103.024529,PhysRevB.105.184509,Huang2022,PhysRevResearch.5.013162,Li_2022,PhysRevB.101.184506,PhysRevB.103.014501,Phillips_2020,arXiv:2105.15205,PhysRevB.106.155119,arXiv:2303.00926,arXiv:2105.15205}.Thanks to its infinite-range interaction, the BHK model can be diagonalized in momentum space, revealing the emergence of the BM state for any finite interaction strength, in contrast to the SF state.
The identification of the BM state is corroborated by several distinctive characteristics, including a unique momentum distribution function, a finite Drude weight, and a vanishing SF weight.
The BM state in the BHK model exhibits several properties reminiscent of a conventional Fermi liquid, such as the Bose surface, a linear temperature-dependent heat capacity, and the saturation of charge susceptibility at low temperatures.
The transition from the MI to the BM phase, driven by band filling, belongs to the universality class of the Lifshitz transition, commonly observed in non-interacting fermionic systems.
It warrants emphasis that the SF state only occurs in the non-interacting limit, indicating that the on-site interaction in momentum space (or infinite-range interaction in real space) is sufficient and crucial for the formation of the BM state.
Importantly, our identified BM state does not rely on the presence of disorder, external magnetic field, or fine-tuning of carrier density. Therefore, the BM state could be considered as a new fixed point of interacting Bose systems.

The subsequent sections of this paper are structured as follows.
Section~\ref{sec1:level1} serves as an introduction to the BHK model, highlighting the key observables employed in the investigation of this model, including the single-particle spectral function $A(k,\omega)$, the Drude weight, the SF {weight}, and the charge susceptibility.
Section~\ref{sec3:level1} presents the results obtained in this study, encompassing the properties of the phase diagram, the MI state, the BM state, and the MI-BM Lifshitz transition at zero temperature.
Section~\ref{sec4:level1} delves into the finite-temperature properties, focusing on the heat capacity and charge susceptibility.
Section~\ref{sec5:level1} further explores the band structure and compressibility, drawing comparisons with the SF state observed in the BH model. Additionally, the resemblance between the BM state and the Fermi liquid is elucidated in the limit of $U \rightarrow \infty$. Finally, Section~\ref{sec6:level1} provides a summary of the findings of this paper.

\section{\label{sec1:level1}the model}	
We consider the following BHK model, which is defined for interacting bosons on a lattice,
\begin{equation}
	\begin{aligned}
	\hat{H}=&-\sum_{i,j}t_{ij}\hat{c}_{i}^{\dag}\hat{c}_{j}-\mu\sum_{j}\hat{c}_{j}^{\dag}\hat{c}_{j}\\
	&+\frac{U}{2N_s}\sum_{j_1,j_2,j_3,j_4}\delta_{j_1+j_3=j_2+j_4}\hat{c}_{j_1}^{\dag}\hat{c}_{j_3}^{\dag}\hat{c}_{j_2}\hat{c}_{j_4}.
	\end{aligned}
\label{eq:HK_model0}
\end{equation}
 Here, $\hat{c}_{j}^{\dag}$ is the creation operator of boson at site $j$ and it satisfies commutative relation $[\hat{c}_{i},\hat{c}_{j}^{\dag}]=\delta_{ij}$. $t_{ij}$ denotes the hoping integral between $i,j$ sites and has translation invariance, i.e. $t_{ij}=t_{i-j}$. We consider a grand-canonical ensemble and the number of bosons can be tuned by varying chemical potential $\mu$.
$N_s$ is the number of lattice sites. The last term in Hamiltonian is the HK interaction, which is infinite-ranged between any four bosons but preserves the center of motion (embodied by the constraint of the $\delta$-function). It should be emphasized that all nontrivial physics come from this interaction since it stabilize a new fixed point.\cite{Huang2022}

Since the HK interaction is local in momentum space, a Fourier transformation on the original Hamiltonian Eq.~\ref{eq:HK_model0} leads to,
\begin{equation} \hat{H}=\sum_k\hat{H}_k=\sum_{k}\left[\epsilon_{k}\hat{n}_{k}+\frac{U}{2}\hat{n}_{k}(\hat{n}_{k}-1)-\mu\hat{n}_{k}\right],
	\label{eq:HK_model1}
\end{equation}
where $\hat{n}_{k}=\hat{c}_{k}^{\dag}\hat{c}_{k}$ is the number of bosons in a state labelled by momentum $k$. It is interesting to note that since $[\hat{H}_{k},\hat{H}_{k}']=0$ for any $k,k'$, the BHK model is a frustration-free model but can have nontrivial physics if each sector $\hat{H}_{k}$ is not trivial \cite{Wen2019}. For simplicity, we consider our system is on hypercubic lattice with only nearest-neighbor-hoping $t$, so the dispersion of bosons is $\epsilon_{k}=-2t\sum_{i=1}^{d}\cos k_{i}$.

To solve Eq.~\ref{eq:HK_model1}, we observe that in each $\hat{H}_k$, $\hat{n}_{k}$ is a good quantum number, thus if we choose $\hat{n}_{k}$'s eigenstate $|n_{k}\rangle$ ($n_k=0,1,2,...$) as basis, $\hat{H}_k$ is automatically diagonalized with its eigen-energy
\begin{equation}
	 E_{n_k}=(\epsilon_{k}-\mu)n_k+\frac{U}{2}n_k({n}_{k}-1).
	\label{eq:energy}
\end{equation}
Particularly, the ground state of $\hat{H}_{k}$ is determined by minimizing $E_{n_k}$, i.e. $\frac{\partial E_{n_{k}}}{\partial n_{k}}=0$, which gives $n_{k}=\mathrm{int} \left[\frac{1}{2}+\frac{\mu-\varepsilon_{k}}{U}\right]$ ($\mathrm{int}[x]$ gives the integer nearest to $x$). For the whole Hamiltonian $\hat{H}$, its eigenstate is just the product-state of each $|n_{k}\rangle$,
 \begin{equation}
 	|n_{k_1},n_{k_2},...,n_{k_{N_s}}\rangle
 	  \equiv (\hat{c}_{{k_1}}^{\dag})^{n_{k_1}}(\hat{c}_{k_2}^{\dag})^{n_{k_2}}...(\hat{c}_{k_{N_s}}^{\dag})^{n_{k_{N_s}}}|0,0,0...\rangle.
 	\label{eq:basis}
 \end{equation}
Thus, without much effort, we have obtained all eigenstates of BHK model, which is a key feature of HK-like models.

Most importantly, the ground state of BHK model can be succinctly expressed as
 \begin{equation}
	|\Psi_g \rangle
	= \prod_{k \in \Omega_0} |0 \rangle_k \prod_{k \in \Omega_1} |1 \rangle_k ...\prod_{k \in \Omega_n} |n \rangle_k,
	\label{eq:gs}
\end{equation}
where $\Omega_n$ represents the momentum space regions with an occupancy of $\langle \hat{n}_{k}\rangle=n$ ($n=0,1,2,3...$). In the subsequent sections of this study, we will be guided by this simple ground-state wavefunction and employ various physical observables, such as the particle distribution function, single-particle spectrum function, Drude weight, SF weight, and charge susceptibility, to construct the phase diagram of the BHK model.
% i.e., $\frac{\partial E_{n_k}}{\partial n_k}=0$

\subsection{\label{sec2:level2} Single-particle Green's function and spectral function}
Firstly, let us define the single-particle (boson) retarded Green's function
\begin{equation}
G^R(t,k)=-i\theta(t)\langle [\hat{c}_{k}(t),\hat{c}^{\dagger}_{k}] \rangle=-i\theta(t)\frac{\mathrm{Tr} \left(e^{-\beta \hat{H}}[\hat{c}_{k}(t),\hat{c}^{\dagger}_{k}]\right)}{\mathcal{Z}}. \nonumber
\end{equation}
Here, $\theta(t)$ is the unitstep function with $\theta(t)=1$ for $t>0$ and vanishes for $t<0$. $\mathcal{Z}=\mathrm{Tr} e^{-\beta \hat{H}}=\prod_{k}\mathcal{Z}_{k}=\prod_{k}\sum_{n_{k}=0}^{\infty}e^{-\beta E_{n_{k}}}$ is the partition function.
We note that in contrast to the case in standard HK model for fermion, here, the summation over $n_{k}$ cannot be performed analytically and numerical calculation with cutoff (define a maximum for $n_{k}$) has to be used. Then, armed with eigenstate Eq.~\ref{eq:basis}, eigen-energy Eq.~\ref{eq:energy}, we have derived the retarded Green's function $G^{R}(\omega,k)$ in terms of the Lehmann spectral representation,
\begin{equation}
	\begin{aligned}	
		&G^R(\omega,k)=\int_{-\infty}^ {\infty} dt e^{i (\omega +i0^{+})t }G^{R}(t,k) \\
		&= \sum\limits_{n_{k},m_{k}} \frac{e^{-\beta E_{n_{k}}}}{\mathcal{Z}_{k}} \left[\frac{|\langle n_{k} | \hat{c}_{k} | m_{k}\rangle|^{2}}{\omega +i0^{+}+E_{n_{k}}-E_{m_{k}}} -\frac{|\langle n_{k} | \hat{c}^{\dagger}_{k} | m_{k}\rangle|^{2} }{\omega +i0^{+}+E_{m_{k}}-E_{n_{k}}}  \right]\nonumber\\
		&= \sum\limits_{n_{k}} \frac{e^{-\beta E_{n_{k}}}}{\mathcal{Z}_{k}} \left[\frac{n_{k}+1}{\omega +i0^{+}+E_{n_{k}}-E_{n_{k}+1}} -\frac{n_{k}}{\omega +i0^{+}+E_{n_{k}-1}-E_{n_{k}}}  \right].
		\label{eq:Lehmann}
	\end{aligned}	
\end{equation}
Obviously, when $T=0$ only the excitation above the groundstate contributes to the $G^{R}$. Thus, the summation over all $n_{k}$ can be neglected and only the term with the groundstate particle occupation $n_{k}=n_{k}^{0}$ is preserved. So, we find the retarded Green's function reduces to an analytical form
\begin{equation}
	\begin{aligned}	
		G^R(\omega,k)=\frac{n_{k}^{0}+1}{\omega-(\varepsilon_{k}+Un_{k}^{0}-\mu)}
		-\frac{n_{k}^{0}}{\omega-(\varepsilon_{k}+U(n_{k}^{0}-1)-\mu)},
		\label{eq:Lehmann2}
	\end{aligned}	
\end{equation}
which is similar to the counterpart in the fermionic HK model $G^R(\omega,k)=\frac{ 1-n_{k}^{0}}{\omega-(\varepsilon_{k}-\mu)}
+\frac{n_{k}^{0}}{\omega-(\varepsilon_{k}+U-\mu)}$ \cite{Phillips_2020}.
The first (second) term in Eq.~\ref{eq:Lehmann2} describes particle (hole) excitation with excitation energy $\omega_{p}=\varepsilon_{k}+Un_{k}^{0}-\mu$ ($\omega_{h}=\mu-\varepsilon_{k}-U(n_{k}^{0}-1)$).

Then, we can obtain the spectral function $A(k,\omega)$ using the relation $A(k,\omega)=-\frac{1}{\pi}\mathrm{Im}G^{R}(k,\omega+i0^{+})$.
This relation allows us to extract valuable information about the system's single-particle excitations and their energy distribution.
It is important to note that $A(k,\omega)$ is positive for $\omega>0$ and negative for $\omega<0$.  In Figure~\ref{fig:spectral}, we present examples of $A(k,\omega)$, which will be further analyzed later.

%In Fig.~\ref{fig:excitation} we plot the critical surface in different systems.
%The Bosonic HK model has the Bose surface different from the general interacting Bosonic system \cite{PhysRevB.93.121109, PhysRevLett.98.220603},
%which every gapless point is associated with a pair of moving modes (see Fig.~\ref{fig:excitation} (b)).
%In the HK model each gapless point is associated with only one right-moving mode or one left-moving mode.
%\textcolor[rgb]{0.00,1.00,0.00}{[I do not understand these arguments and the Fig.~\ref{fig:excitation}!]}

% \begin{figure}
% 	\centering
% 	\includegraphics[width=1.0\columnwidth]{lowE_excitation.jpg}
% 	\caption{The schematic picture of the low-energy excitation for (a) the Fermi surface; (b) the general BM state;
% 		(c) the BM in the HK model at $W/U=2,\mu/U=0.5$.}
% 	\label{fig:excitation}
% \end{figure}

\subsection{\label{sec2:level2} Drude weight and superfluid weight}
Next, following the general strategy of many-body physics to distinguish metallic and insulating states, we try to calculate the Drude weight and SF weight, which are the most relevant transport quantities.\cite{PhysRevLett.68.2830,Mucio}
The Drude weight $D$ and SF weight $D_{s}$ can be deduced by studying two different limiting behaviors of the current-current correlation function $\chi_{j_{x} j_{x}}(\vec{q},\omega)$, which represents the paramagnetic component of the linear-current-response induced by the vector potential $A_x(\vec{q},\omega)$
\begin{equation}
	\begin{aligned}	
		\langle j_x(\vec{q},\omega)\rangle=-[e^2(\langle-K_x\rangle-\chi_{j_{x} j_{x}}(\vec{q},\omega))A_x(\vec{q},\omega)],
		\label{eq:lcr}
	\end{aligned}
\end{equation}
\begin{equation}
	\begin{aligned}	
		\chi_{j_{x} j_{x}}(\vec{q},\omega)\equiv \frac{i}{N_s}\int_{-\infty}^{\infty} dt \theta (t) \langle \left[ j_x^p(\vec{q},t), j_x^p(\vec{-q},0)\right] \rangle e^{i \omega t}.
		\label{eq:Drude}
	\end{aligned}
\end{equation}
The first term of Eq.~\ref{eq:lcr} is the diamagnetic term, which contributes from the kinetic energy per site divided by the number of dimensions, i.e. $K_{x}=-\frac{t}{N_{s}}\sum_{j}(c_{j+x}^{\dag}c_{j}+c_{j}^{\dag}c_{j+x})$. The paramagnetic current is defined by $j_{x}^{p}(\vec{q})=-{it}\sum_{j}e^{-i\vec{q}\cdot \vec{R}_{j}}(c_{j}^{\dag}c_{j+x}-c_{j+x}^{\dag}c_{j})$.

%$\Lambda_{xx}(\vec{q},\omega)$ in the second term is the current-current correlation function, which is deducted from the paramagnetic term.
%The uniform frequency-dependent conductivity $\sigma_{xx}(\omega)$ refers to the current response to the electric field,
%$E_x(\vec{q=0},\omega)=i\omega A_x(\vec{q=0},\omega)$.
The Drude weight is given by the $\delta$-function part of the uniform conductivity $\sigma_{xx}(\omega)\equiv-e^{2}\frac{\langle-K_{x}\rangle-\chi_{j_{x}j_{x}}(\vec{q=0},\omega)}{i\omega}$ as $\omega \rightarrow 0$,
\begin{equation}
	\begin{aligned}	
		\frac{D}{\pi e^2}=\langle-K_x\rangle-\chi_{j_{x} j_{x}}(\vec{q=0},\omega \rightarrow 0).
		\label{eq:Drude1}
	\end{aligned}
\end{equation}
If the order in which $\vec{q}$ and $\omega$ approach zeros is exchanged, one obtains the SF weight
\begin{equation}
	\begin{aligned}	
		\frac{D_s}{\pi}=\langle-K_x\rangle-\chi_{j_{x} j_{x}}(\vec{q} \rightarrow 0, \omega = 0).
		\label{eq:Drude2}
	\end{aligned}
\end{equation}

\renewcommand\arraystretch{1.2}
\begin{table}
	
	\fontsize{8}{20}
	\setlength{\tabcolsep}{6mm}
	\begin{tabular}{c|c|c|c}
		\toprule
		\toprule
		\diagbox [width=8em,trim=l]{variable}{state} &BM & MI & SF \\
		\hline
		$D_s$& 0&0& $\neq0$ \\
		\hline
		D& $\neq$ 0& 0& $\neq0$\\
		\bottomrule
		\bottomrule
	\end{tabular}
	\caption{The Drude weight and SF weight for different states.}
	\label{tb:1}
\end{table}

%According to Ref.~\cite{Coleman},
%$\frac{D}{\pi}$ is the so-called London response kernel, thus the paramagnetic part can be written in the real time
% \begin{equation}
%	\begin{aligned}	
%		\frac{D}{\pi e^2}=\langle-k_x\rangle- \chi_{j_{x} j_{x}}(\vec{q=0},\omega \rightarrow 0),
%		\label{eq:Drude3}
%	\end{aligned}
%\end{equation}
%where  \textcolor[rgb]{0.00,1.00,0.00}{[Why you use the symbol $\chi_{j_{x} j_{x}}$ instead of $\Lambda_{xx}$?]}

Here we assume $q_x=0$ for the $\vec{q} \rightarrow \vec{0}$ limit, since the London gauge requires that $\vec{q} \cdot \vec{A}=0$.
The current-current correlation function in the $q_x=0$ situation can be written in a compact way
\begin{equation}
	\begin{aligned}		
		&\chi_{j_{x} j_{x}}(\vec{q},\omega)= \frac{i}{N_s}\int_{-\infty}^{\infty} dt e^{i\omega t} \theta (t) \langle [j_{x}(\vec{q},t),j_{x}(\vec{-q},0)]\rangle \\
		&=-i\frac{t^2}{N_s}\sum_{k_1,k_2}(e^{-i2k_x}+e^{i2k_x}-2) \times \\
		&\int_{-\infty}^{\infty} dt e^{i\omega t} \theta (t)
		\left\langle\left[ c_{\vec{k_1}}^{\dag}(t)c_{\vec{k_1+q}}(t),c_{\vec{k_2}}^{\dag}(0)c_{\vec{k_2-q}}(0)\right]\right\rangle \\
		& =\frac{t^2}{N_s}\sum_{\vec{k_1},\vec{k_2}}(e^{-i2k_x}+e^{i2k_x}-2) \ll c_{\vec{k_1}}^{\dag}c_{\vec{k_1+q}}\mid c_{\vec{k_2}}^{\dag}c_{\vec{k_2-q}} \gg_{\omega},
		\label{eq:eq8}
	\end{aligned}
\end{equation}
where $\ll c_{\vec{k_1}}^{\dag}c_{\vec{k_1+q}}\mid c_{\vec{k_2}}^{\dag}c_{\vec{k_2-q}} \gg_{\omega}$ is the retarded Green's function in real frequency domain.

We first focus on the Drude weight (Eq.~\ref{eq:Drude1}). The current-current correlation function now is simplified as
\begin{equation}
	\begin{aligned}		
		&\chi_{j_{x} j_{x}}(\vec{q=0},\omega \rightarrow 0)\\
		&=\frac{t^2}{N_s}\sum_{\vec{k_1},\vec{k_2}}(e^{-i2k_x}+e^{i2k_x}-2) \ll c_{\vec{k_1}}^{\dag}c_{\vec{k_1}}\mid c_{\vec{k_2}}^{\dag}c_{\vec{k_2}} \gg_{\omega},
		\label{eq:Drude4}
	\end{aligned}
\end{equation}
Since the scattering process (or the interaction between bosons) in the BHK model preserves the momentum, $\hat{n}_{k}$ is a good quantum number, leading to $\ll c_{\vec{k_1}}^{\dag}c_{\vec{k_1}}\mid c_{\vec{k_2}}^{\dag}c_{\vec{k_2}} \gg_{\omega}=0$.
Therefore, for any $\mu/U$ and $W/U$, the current-current correlation function is invariant as zero.
The Drude weight satisfies $\frac{D}{\pi e^2}=\langle-K_x\rangle$. Therefore, the bosons transport without dissipation, and the conductivity is completely determined by the average kinetic energy.

Next we consider the SF weight in the opposite limitation.
In the BHK model, the effect of interaction has been accounted in the non-trivial distribution function $\langle\hat{n}_{k}\rangle$, while eigenstates retain a Fock-like form. Thus, the current-current correlation function can be expressed as
\begin{equation}
	\begin{aligned}	
		&\chi_{j_{x} j_{x}}(\vec{q }\rightarrow \vec{0},\omega = 0)\\
		&=\frac{t^2}{N_s}\sum_{\vec{k_1},\vec{k_2}}(e^{-i2k_x}+e^{i2k_x}-2)\ll c_{\vec{k_1}}^{\dag}c_{\vec{k_1+q}}\mid c_{\vec{k_2}}^{\dag}c_{\vec{k_2-q}} \gg_{\omega}\\
		&=\frac{t^2}{N_s}\sum_{\vec{k_1},\vec{k_2}}(e^{-i2k_x}+e^{i2k_x}-2)\delta_{\vec{k_2}, \vec{k_1+q}}\frac{n_{\vec{k_1+q} }-n_{\vec{k_1}}}
		{\omega -\epsilon_{\vec{k_1+q}}+\epsilon_{\vec{k_1}}}.
		\label{eq:eq10}
	\end{aligned}
\end{equation}

For metallic states, the Drude weight $D$ must be nonzero while insulators have $D=0$. Furthermore, to identify SF states from generic metallic states, one expects that SF states with SF {weight} $D_{s}\neq0$ and $D\neq0$. (See also Table~\ref{tb:1}) Based on this prescription, in this work, we will find that BHK model have BM and MI states but no SF. (Fig.~\ref{fig:SF})
\subsection{\label{sec2:level2} Charge susceptibility}
In addition to directly calculating the partial derivative of the particle number with respect to the chemical potential, the charge susceptibility $\chi_c$ can also be obtained through the density-density correlation function. This correlation function, denoted as $\chi_c(R_i,R_j,t)$,  is defined as the time-ordered commutator between the particle density operator $\hat{n}_{i}$ at site $R_i$ and the particle density operator $\hat{n}_{j}$ at site $R_j$. It can be expressed as follows:
\begin{equation}
	\begin{aligned}	
		\chi_c(R_i,R_j,t)=&-i\theta(t)\langle \left[ \hat{n}_i(t),\hat{n}_j\right] \rangle\\
		=&\frac{-i}{N_s^2}\sum_{{k_1},{k_2},{k_3},{k_4}}e^{-i({k_1}-{k_2})R_i}e^{-i({k_3}-{k_4})R_j}
		 \times \\
		&\theta(t)\left\langle \left[ \hat{c}^{\dagger}_{{k_1}}(t)\hat{c}_{{k_2}} (t),\hat{c}^{\dagger}_{{k_3}}\hat{c}_{{k_4}} \right]\right\rangle.
		\label{eq:eq10}
	\end{aligned}
\end{equation}
This expression can be further transformed into momentum and frequency space using Fourier transformation:
\begin{equation}
	\begin{aligned}	
		&\chi_c({q},\omega)=\frac{1}{N_s} \sum_{R_i,R_j}e^{-iq(R_i-R_j)} \int^{\infty}_{0}dt e^{i\omega t} \chi_c(R_i,R_j,t) \\
		&=\frac{1}{N_s} \sum_{{k_1},{k_3}}\ll c_{{k_1}}^{\dag}c_{{k_1+q}}\mid c_{{k_3}}^{\dag}c_{{k_3-q}} \gg_{\omega}.
		\label{eq:eq10}
	\end{aligned}
\end{equation}
The (static and uniform) charge susceptibility, which serves as an indicator of the phase boundary, is defined as the retarded Green's function at the zero-momentum and zero-frequency limit, i.e., $\chi_c=\chi_c({q=0},\omega=0)=\frac{1}{N_{s}}\frac{\partial N}{\partial\mu}$.

\section{\label{sec3:level1} results}

\subsection{\label{sec2:level2} The ground-state phase diagram}	
\begin{figure}
	\centering
	\includegraphics[width=1\columnwidth]{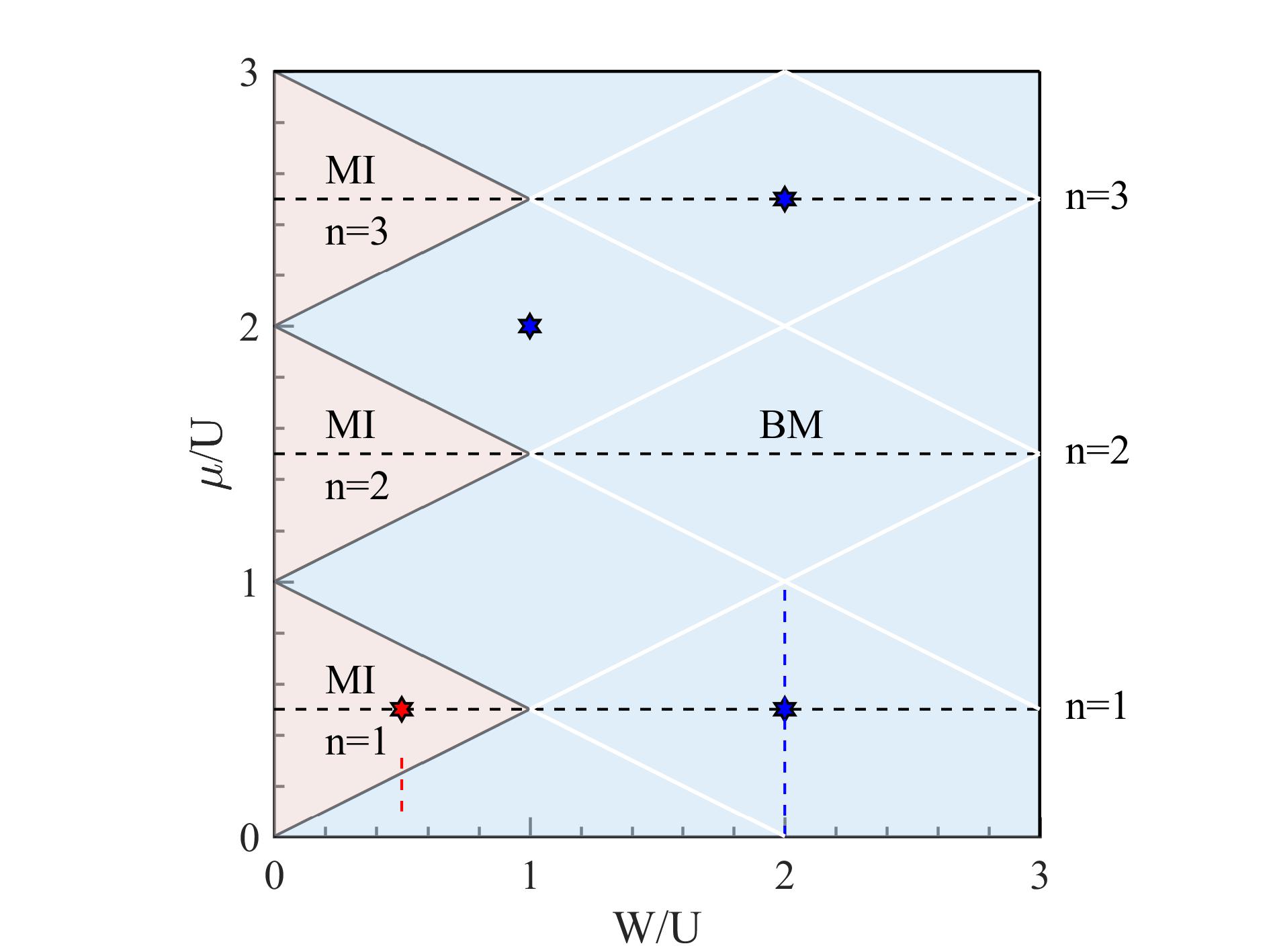}
	\caption{The zero-temperature phase diagram of the Bose Hastugai-Khonmoto (BHK) model for any dimension. The phase diagram consists of the Mott insulator (MI) region (shown in red) at small $W/U$ and the Bose metal (BM) region (shown in blue) at large $W/U$. Three red triangle regimes correspond to the MI states with different boson densities $n$, where $n=1,2,3$ from bottom to top.
		The black line represents the Lifshitz transition between the MI and BM phases, while the white line denotes the Lifshitz transition between different BM states.
		The black dashed lines represent the BM state with an invariant integer boson density.
		The specific MI/BM state we focus in the latter of this paper is denoted by red/blue hexagon.
		%	The blue (red) hexagon denotes the MI (BM) states we most focus in this paper.
		The scaling behavior of the BM-MI phase transition along the red dashed line is demonstrated in Fig.~\ref{fig:scaling}, while the thermodynamic properties of the BM state along the blue dashed line are shown in Fig.~\ref{fig:Cv}.}
	\label{fig:phasediagram}
\end{figure}

Before delving into the intricacies of our calculations, we present our key finding encapsulated in the zero-temperature phase diagram
(plotted on the $\mu-W$ plane), applicable to any spatial dimension (see Fig.~\ref{fig:phasediagram}). Here, $W=4td$ represents the bandwidth of a hypercubic lattice with nearest-neighbor hopping.
Our analysis reveals that the BHK system is predominantly governed by two distinct states: the MI state characterized by an integer density, and the BM state exhibiting varying densities.
We emphasize that the MI state exhibits the greatest stability at each $\mu/U=\frac{2n-1}{2}$ (where
$n$ is an integer denoting the number density), wherein the boson density remains invariant with changing bandwidth. In this situation, the fixed-density MI-BM transition (or interaction-driven MI-BM transition) occurs at $U_c=W$.
In the subsequent sections, we focus primarily on discussing specific parameter values, including one MI state ($W/U=0.5,  \mu/U=0.5$) and three BM states ($W/U=2, \mu/U=0.5$, $W/U=1, \mu/U=2$, and $W/U=2, \mu/U=2.5$).
For visual guidance, these specific points are marked with hexagonal symbols in Fig.~\ref{fig:phasediagram}.

We would like to emphasize a significant departure from the original study of the BHK model\cite{CONTINENTINO1994619}, wherein our findings reveal a remarkable substitution of the SF state with an unexpected BM state for any finite interaction strength $U$. Consequently, in the case of integer boson filling, our results demonstrate an MI-BM transition instead of the anticipated MI-SF transition as the interaction strength is increased.
The precise reasons for the erroneous identification of the SF state by the authors in Ref.~\onlinecite{CONTINENTINO1994619} remain elusive to us. However, it is worth noting that their study lacks any discernible calculations pertaining to charge susceptibility, SF weight, and Drude weight.

\begin{figure}
	\centering
	\includegraphics[width=0.9\columnwidth]{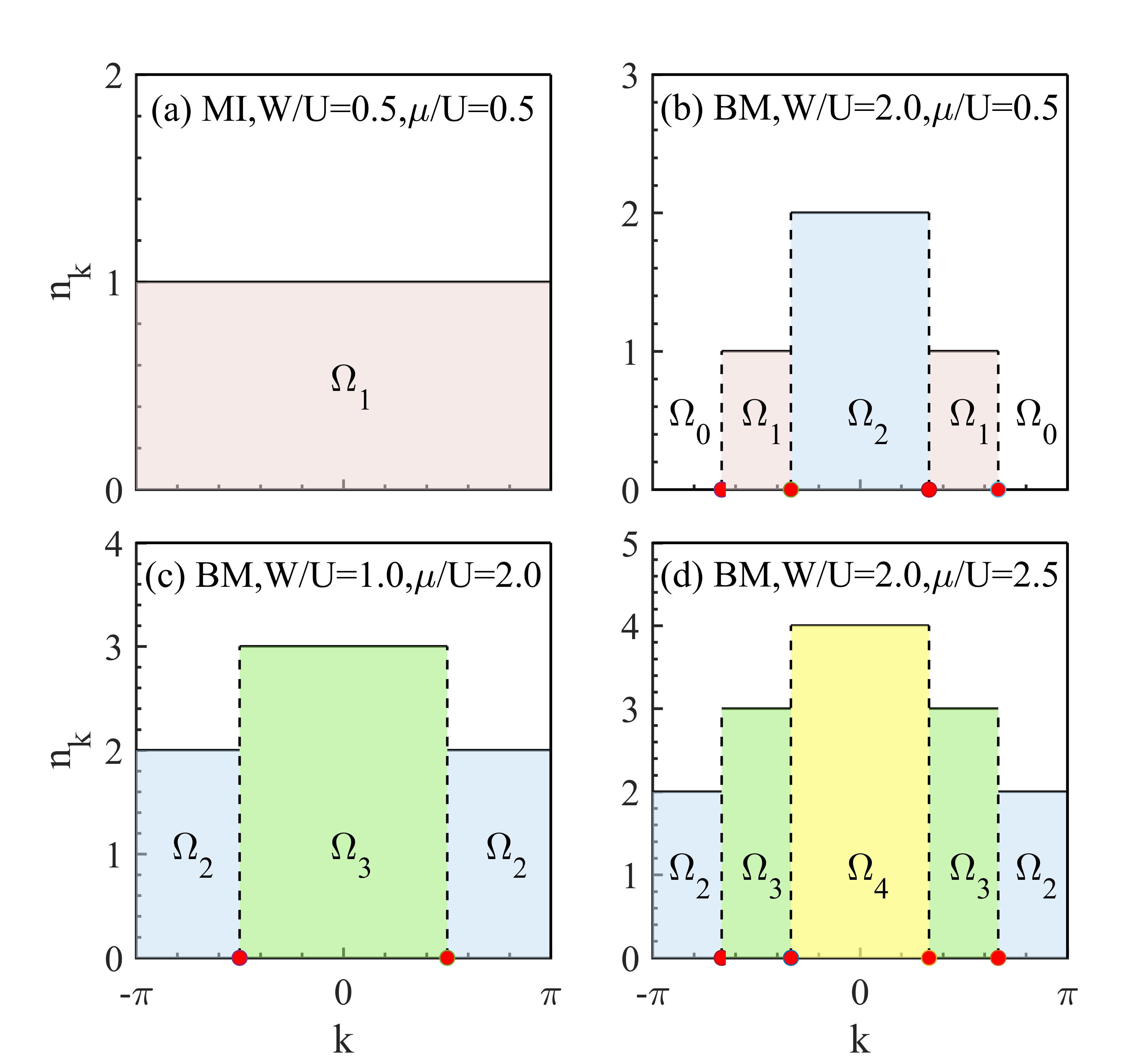}
	\caption{Boson's distribution function $\langle\hat{n}_k\rangle$ in the one-dimensional BHK model for (a) the MI $W/U=0.5$, $\mu/U=0.5$; (b)the BM $W/U=2$, $\mu/U=0.5$;
		(c) the BM $W/U=1$, $\mu/U=2$; (d) the BM $W/U=2$, $\mu/U=2.5$.
		The occupation number $n$ in different momentum regimes is denoted by $\Omega_n$.
		The Bose surface (gapless point) is denoted by the red circle, located at the step of the distribution function.}
	\label{fig:dos2}
\end{figure}

\begin{figure}
	\centering
	\includegraphics[width=1\columnwidth]{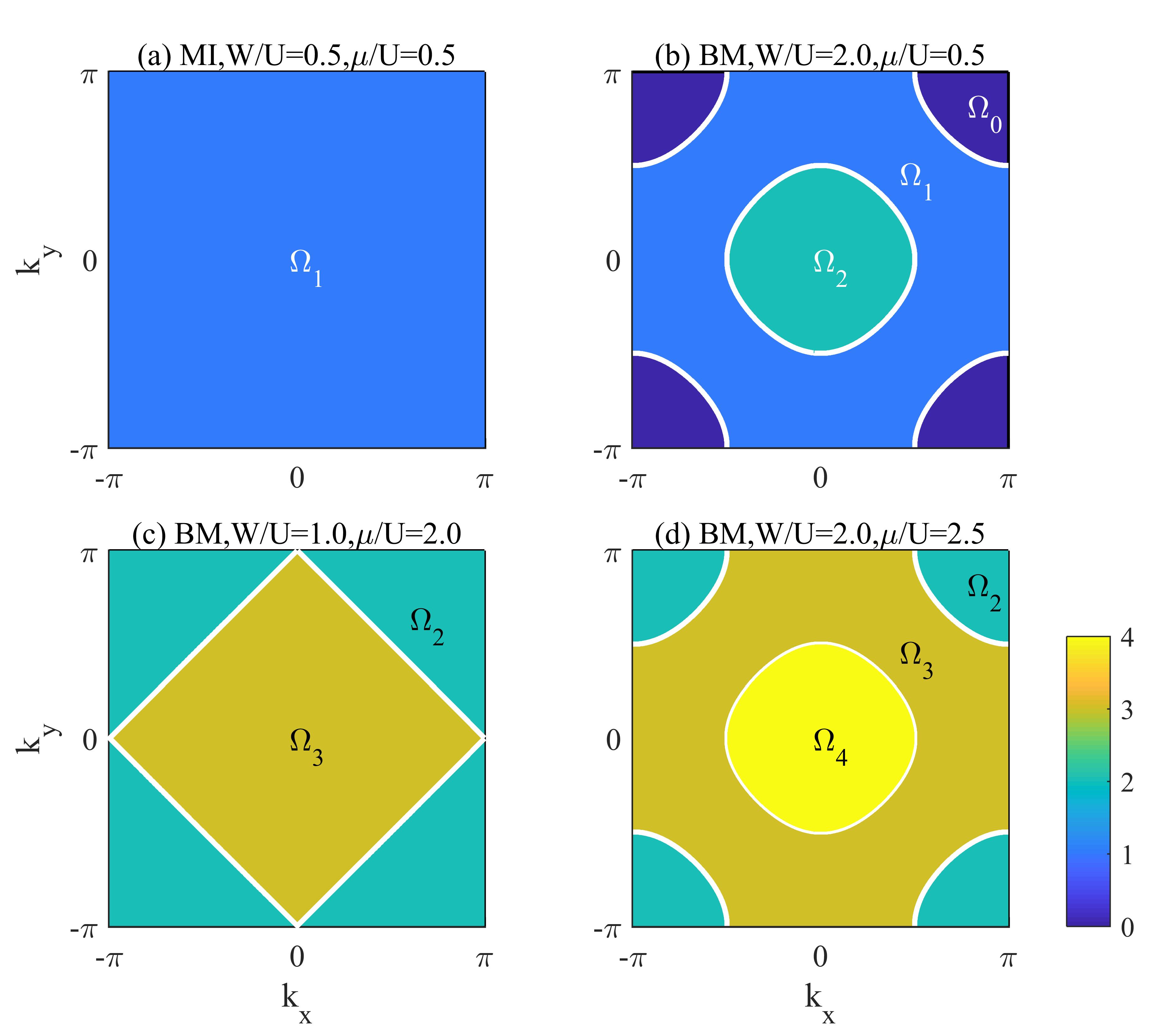}
	\caption{Boson's distribution function $\langle\hat{n}_k\rangle$ in the two-dimensional HK model for (a) the MI $W/U=0.5$, $\mu/U=0.5$; (b)the BM $W/U=2$, $\mu/U=0.5$;
		(c) the BM $W/U=1$, $\mu/U=2$; (d) the BM $W/U=2$, $\mu/U=2.5$.
		The Bose surface is denoted by write circle. Zero-energy excitation is located at the intersection of different $\Omega_n$ region. }
	\label{fig:Aw}
\end{figure}

\begin{figure*}
	\centering
	\includegraphics[width=1.8\columnwidth]{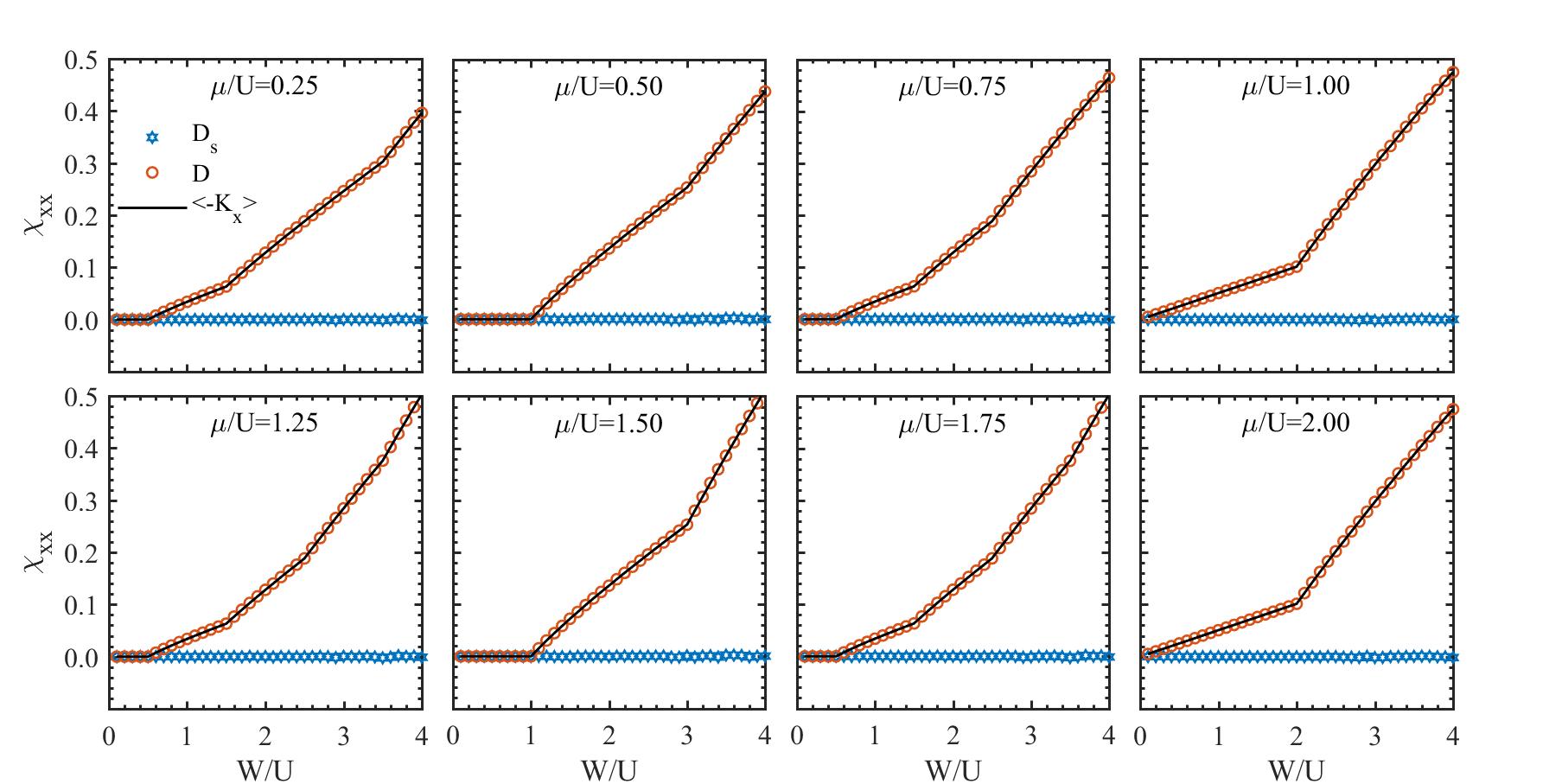}
	\caption{The superfluid weight $D_s$ (blue hexagon) and the Drude weight $D$ (red star) in the two-dimensional BHK model for different chemical potential $\mu/U$ with varying $W/U$. The black line denotes minus kinetic energy per site divided by lattice dimension.
		The superfluid weight is invariant zero for both MI and BM states. The Drude weight is equal to $\langle -k_x \rangle$, indicating that no resistivity is caused by the HK interaction.}
	\label{fig:SF}
\end{figure*}
\begin{figure}
	\centering
	\includegraphics[width=0.9\columnwidth]{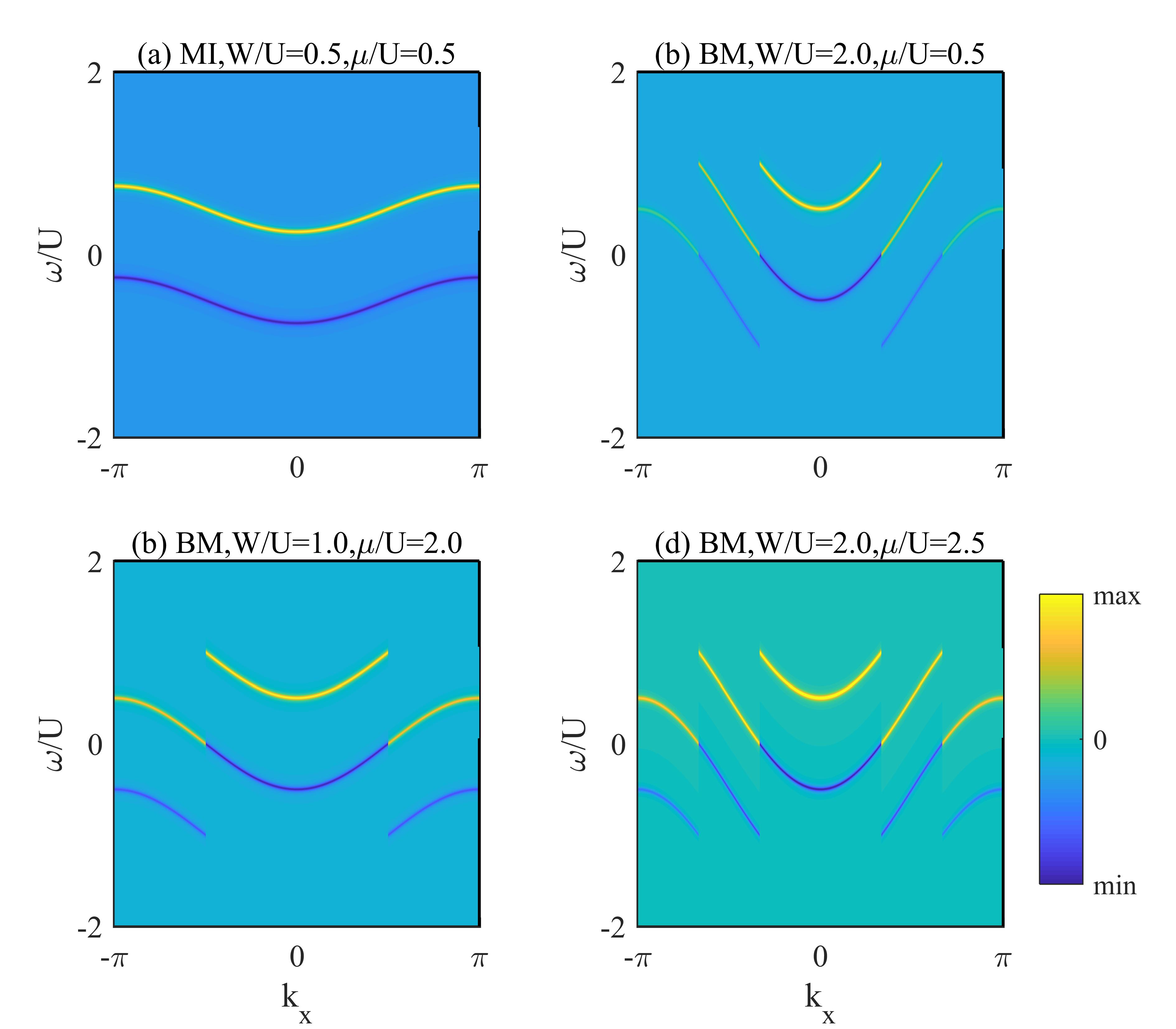}
	\caption{The spectral function of the BHK model for (a) the MI $W/U=0.5$, $\mu/U=0.5$; (b)the BM $W/U=2$, $\mu/U=0.5$;
		(c) the BM $W/U=1$, $\mu/U=2$; (d) the BM $W/U=2$, $\mu/U=2.5$.
	}
	\label{fig:spectral}
\end{figure}
\begin{figure}
	\centering
	\includegraphics[width=1.1\columnwidth]{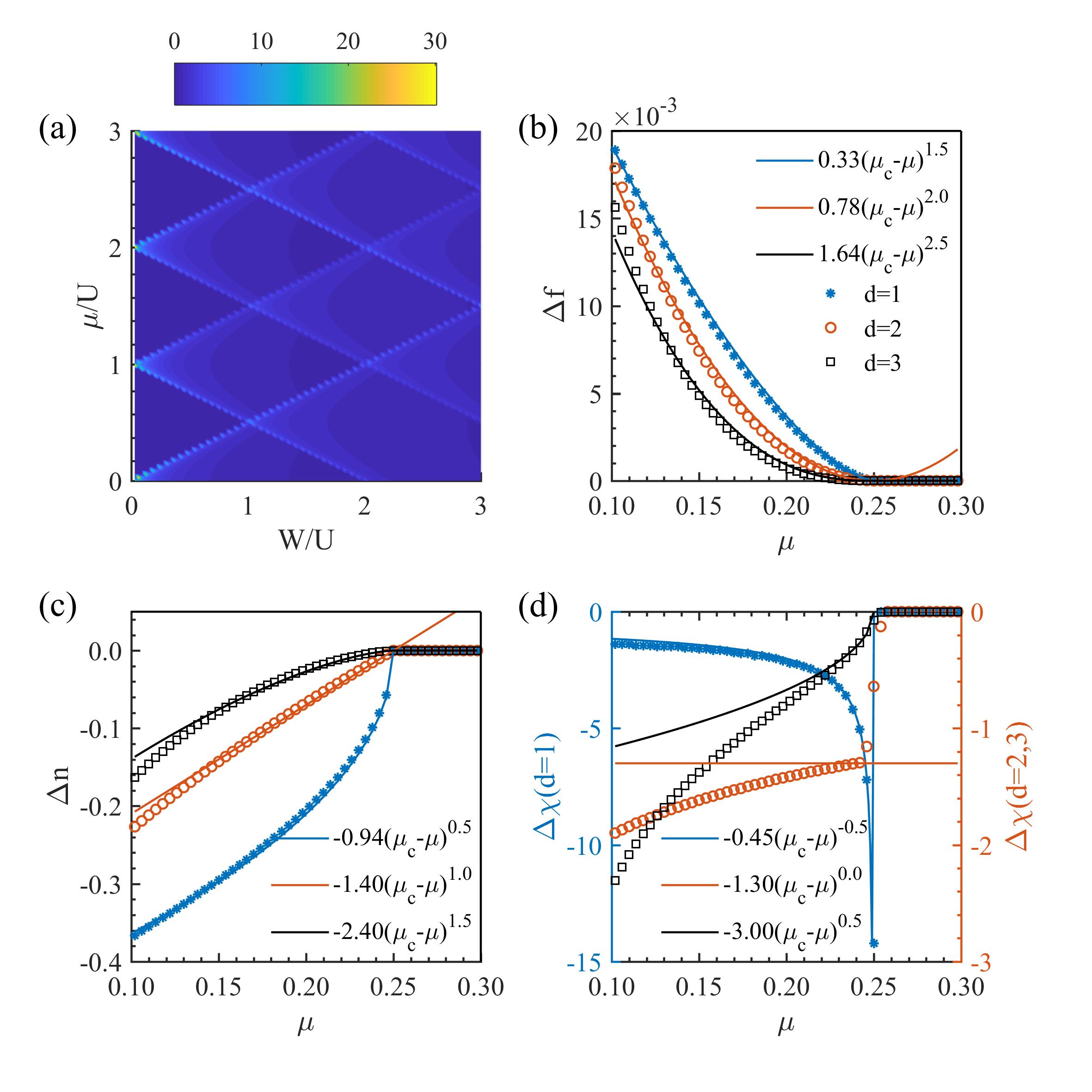}
	\caption{(a) The charge susceptibility $\chi_c$ evaluated on the $W$-$\mu$ plane for a $d=1$ BHK system. The calculations are performed on a lattice with $N_s=40000$ sites, and the obtained results are qualitatively consistent with those expected in higher-dimensional scenarios. (b) Free energy density, (b) particle density, and (c) charge susceptibility around critical point $\mu_c=0.25$ versus fitted scaling formula Eq.~\ref{eq:scale} with $W/U=0.5$. }
	\label{fig:scaling}
\end{figure}
\subsection{\label{sec2:level2} Mott insulator}	
The MI state here occurs when the number of bosons $N$ is commensurate with the number of lattice sites $N_{s}$, i.e. $n=N/N_{s}=1,2,3...$. Guided by Eq.~\ref{eq:gs}, it is clear that the wavefunction of the MI state is given by:
\begin{equation}
 |\Psi_\text{Mott} \rangle= \prod_{k \in \text{BZ}} |n \rangle_k,
\end{equation}
where all momentum states are occupied by the same number of bosons, denoted as $n$. This argument is supported by the distribution function of bosons $\langle\hat{n}_k\rangle$, shown in Figs.~\ref{fig:dos2} (a) and \ref{fig:Aw} (a), for the $1D$ chain and the $2D$ square lattice, respectively.
The particle and hole excitation above $|\Psi_\text{Mott} \rangle$ can be constructed as $\hat{c}_{q}^{\dag}|\Psi_\text{Mott} \rangle,\hat{c}_{q}|\Psi_\text{Mott} \rangle$, whose excitation energy are denoted as
$\omega_{p}=\varepsilon_{q}+Un-\mu$, $\omega_{h}=-\varepsilon_{q}-U(n-1)+\mu$.

The stability of MI requires $\omega_{p},\omega_{h}>0$ for any momentum $q$, thus
we can establish the MI regime in the ground state, which has been plotted in Fig.~\ref{fig:phasediagram}. To be specific, for given $n$,
we have $\omega_{p}=\varepsilon_{k}+Un-\mu,\omega_{h}=\mu-\varepsilon_{k}-U(n-1)$. Then, $\omega_{p}\rightarrow 0^{+}$ gives
\begin{equation}
\frac{\mu}{U}=\frac{(\varepsilon_{k})_{min}}{U}+n,\nonumber
\end{equation}
while $\omega_{h}\rightarrow 0^{+}$ gives
\begin{equation}
\frac{\mu}{U}=\frac{(\varepsilon_{k})_{max}}{U}+n-1.\nonumber
\end{equation}
Here, $(\varepsilon_{k})_{min}$ and $(\varepsilon_{k})_{max}$ refer to the energy of the band bottom and top of the free bosons, respectively. Since $(\varepsilon_{k})_{min}=-W/2$ and $(\varepsilon_{k})_{max}=W/2$. The above two equations plus $W/U=0$ axis give us the regime of MI as shown in Fig.~\ref{fig:phasediagram}.

Because, $\omega_{p},\omega_{h}>0$ in MI, the charge susceptibility has to be vanished, i.e. $\chi_{c}=\frac{1}{N_{s}}\frac{\partial N}{\partial\mu}=0$, which is the key signature of the insulating nature of MI. With the same reason, both Drude weight $D$ and SF weight $D_{s}$ are zero in MI. (Fig.~\ref{fig:SF})

Furthermore, as evident from the spectral function $A(\omega,k)$ at zero temperature, Fig.~\ref{fig:spectral}(a) with momenta chosen along the path from $(-\pi,-\pi...-\pi)$ to $(0,0,..0)$ and to $(\pi,\pi...\pi)$, the MI state is characterized by a full-filled band (the dominated negative weight of $A(\omega,k)$).
For any given $k$, a finite energy is required for a boson to be excited or removed, which is consistent with the analysis of the wavefunction and the requirement of $\omega_{p},\omega_{h}>0$.

\subsection{\label{sec2:level2} Bose metal}
We have seen that the stability of MI leads to $\omega_{p},\omega_{h}>0$, otherwise, the gap for the particle and/or hole excitation can be vanished and MI must break down in this case.
Our objective is to investigate the potential states that emerge when the breakdown of the MI occurs.
%The aim of us is to explore which states are the candidate when MI breaks down.

For those familiar with the BH model and Bogoliubov's SF theory, the SF state emerges as a highly plausible candidate in this context. It is well-established that in the SF state, bosons tend to condense into a single momentum point, i.e. the condensation momentum $k_{0}$. For the dispersion $\epsilon_{k}=-2t\sum_{i=1}^{d}\cos k_{i}$, we find that $k_{0}=(0,0,...0)$, i.e. the bottom of $\epsilon_{k}$. The condensation in $\mid k_0 \rangle$ yields a SF wave-function like $(\hat{c}_{k_{0}}^{+})^{N}|0,0,0...\rangle=|N\rangle_{k_{0}}$.
However, in contrast to SF, bosons in the BHK model do not condense into $k_{0}$ owing to the energy penalty from HK interactions. This significant difference between the SF state and our calculations is depicted in Figs.~\ref{fig:dos2}(b-d) and \ref{fig:Aw}(b-d). Analogous to the Fermi surfaces in Fermi liquid/gas, distinct surfaces separate regimes with different particle numbers ($\Omega_n$).
We refer to these gapless surfaces as the 'Bose surfaces', and to the states with Bose surfaces as the BM state.

The Bose surfaces live at discrete momentum points (indicated by red solid circles in Fig.~\ref{fig:dos2} (b-d)) in one dimension, and it is more appropriate to term these points as Bose points, similar to their counterpart, i.e., the Fermi point in a one-dimensional Fermi liquid. In the case of a two-dimensional square lattice, the Bose surfaces form closed loops (represented as white lines in Fig.~\ref{fig:Aw} (b-d)), akin to typical Fermi surfaces on a square lattice.

Let us consider a simple example. For Fig.~\ref{fig:dos2}(b), there are two pairs of Bose points and are denoted as $\pm k_{01},\pm k_{12}$ which separates regimes with $\langle\hat{n}_{k}\rangle=0,1$ and $\langle\hat{n}_{k}\rangle=1,2$, respectively. Now, if we consider correlation function or single-particle density matrix $\langle c_{i}^{\dag}c_{j}\rangle$, it is found that
\begin{eqnarray}
	\begin{aligned}
&\langle c_{i}^{\dag}c_{j}\rangle=\frac{1}{N_{s}}\sum_{k}e^{-ik(R_{i}-R_{j})}\langle\hat{n}_{k}\rangle \nonumber \\
&=\left(\int_{-k_{01}}^{-k_{12}}+\int_{k_{12}}^{k_{01}}\right)\frac{dk}{2\pi}e^{-ik(R_{i}-R_{j})}\nonumber
+\int_{-k_{12}}^{k_{12}}\frac{dk}{2\pi}2e^{-ik(R_{i}-R_{j})}\nonumber\\
&=\frac{2}{2\pi}\frac{\sin k_{01}(R_{i}-R_{j})+\sin k_{12}(R_{i}-R_{j})}{R_{i}-R_{j}},
\label{eq:density}
\end{aligned}
\end{eqnarray}
which just like the case of a free fermion system with Fermi wavevector $k_{01},k_{12}$. In other words, if we add a nonmagnetic impurity into BHK model, we will expect that there exists a Friedel oscillation with characteristic wavevector $k_{01}$ and $k_{12}$.\cite{arXiv:2303.00926} This may provide a practical approach to detect the Bose point or the generic Bose surface if they indeed exist. Moreover, when $|R_{i}-R_{j}|\rightarrow\infty$, we see $\langle c_{i}^{\dag}c_{j}\rangle\rightarrow0$ so SF-like long-ranged order for boson does not in the BM state. Such fact is valid for all spatial dimension and for all $U>0$.
In addition, we note that when $R_{i}=R_{j}$, the boson's density is $(2k_{01}+2k_{12})/(2\pi)$, which acts as a Luttinger theorem for BM states \cite{PhysRevB.61.7941}.

To gain further insight into the behavior of the BM, MI or SF in the BHK model, we examine the Drude weight $D$ and SF weight $D_{s}$, as depicted in Fig.~\ref{fig:SF}.\cite{PhysRevLett.68.2830,Coleman}.
In the presence of a non-zero SF and Drude weight ($D_s \neq 0$, $D \neq 0$), an SF state is expected. Conversely, a metal is characterized by a zero SF weight and a non-zero Drude weight ($D_s=0$, $D \neq 0$). When both the SF and Drude weight vanish ($D_s=0$,  $D = 0$), it indicates the presence of an MI state.
Fig.~\ref{fig:SF} presents the variations of $D$ (represented by red stars) and
$D_{s}$ (represented by blue hexagons) for different
$\mu/U$ values with varying bandwidth
$W/U$. As a reference, the diamagnetic response term, i.e., the minus total kinetic energy per dimension
$\langle -K_x \rangle$, is plotted as a black line.
For small values of $W/U$, the MI state is confirmed, as evidenced by the vanishing values of both $D$ and $D_s$.
As $W/U$ increases, the SF weight $D_s$ remains zero, while the Drude weight $D$ grows at some critical $W_c$, which is consistent with the evolution of the structure of the momentum distribution during the MI-BM transition.
The persistent absence of SF weight across different parameter regimes further confirms the absence of the SF state within the finite-$U$ regime.
Associated with the unique distribution function, we conclude that the BM states prevail over the SF state in the metallic regimes of the BHK model.

Furthermore, we present the spectral function of the BM states in Fig.~\ref{fig:spectral}(b-d).
In this context, the yellow (blue) line represents the first (second) term in Eq.~\ref{eq:Lehmann}, which signifies the single-particle (single-hole) excitation.
Within the single-particle spectrum, the Bose surface corresponds to the points where the spectral function undergoes continuous sign changes within a single band.
Similar to the Fermi liquid, the MI and BM can also be distinguished by the absence or presence of a Bose surface.
In the MI state, where the chemical potential resides within the energy gap (as depicted in Fig.~\ref{fig:spectral}(a)), no Bose surface is observed. Conversely, the BM states can exhibit multiple Bose surfaces, as illustrated in Fig.~\ref{fig:spectral}(b-d).

\subsection{\label{sec3:level2}The Lifshitz transition}

To elucidate the putative phase transition in the BHK model, we begin by plotting the global charge susceptibility in the $W$-$\mu$ plane for a one-dimensional system (see Fig.~\ref{fig:scaling} (a)).
The divergent $\chi_c$ delineates the phase boundary between distinct states, which can be expressed as
\begin{equation}
	\begin{aligned}	
	\mu_{c1}=n&U+\frac{W}{2},\\
	\text{or} \\
	\mu_{c2}=n&U-\frac{W}{2},
	\label{eq:scale0}
\end{aligned}	
\end{equation}
where $n=1,2,...$. Notably, $\mu_{c1}$ and $\mu_{c2}$ corresponds precisely to
the the top and bottom of the $n$-th band, respectively.
Based on the location of divergence and the evolution of $\langle\hat{n}_k\rangle$ discussed earlier, we anticipate that these zero-temperature quantum transitions are connected to Lifshitz transitions \cite{Mucio}.
For a $d$-dimensional system with near the Lifshitz transition point, the dynamical critical exponents $z$, the correlation length exponent $\nu$, and the critical exponent $\alpha$ should satisfy $z=2, \nu=1/2, \alpha=1-d/2$, respectively.
In conventional, $g=\mu-\mu_c$ is the natural variable for Lifshitz transitions, signifying the distance of the chemical potential $\mu$ from the band top or bottom.
To verify this conjecture for our BHK model, we study the scaling behavior of the free energy density $f$, the particle density $n$, and the charge susceptibility
$\chi_c$ around the phase transition, which should follow
\begin{equation}
	\begin{aligned}	
&\Delta f=f-f_0 \sim (\mu-\mu_c)^{d/2+1}\\
&\Delta n=n-n_0 \sim (\mu-\mu_c)^{d/2}\\
&\Delta \chi=\chi-\chi_0 \sim (\mu-\mu_c)^{d/2-1}.
		\label{eq:scale}
	\end{aligned}	
\end{equation}
Here, $f_0$, $n_0$, and $\chi_0$ represent certain background values that need to be subtracted.
For free energy density ($f=E-\mu  n$), the chemical potential energy is subtracted to avoid the influence of the linear term $n(\mu-\mu_c) $.
We now focus on the specific case $W/U=0.5$ around $\mu_{c2}/U=0.25$, where the critical chemical potential of the BM-MI transition locates on the top of the lowest band.
As illustrated in Fig.~\ref{fig:scaling} (b-d), we examine the scaling behavior of the quantum phase transition for different dimensions ($d=1,2,3$).
It is evident that in the metallic regime, $f$, $n$, and $\chi_c$ all exhibit behavior consistent with the critical exponents of the Lifshitz transition.
It is worth noting that varying the bandwidth at a fixed chemical potential also induces transitions by changing the variable $\mu-\mu_c$.
Consequently, we conclude that both the chemical potential-driven and bandwidth-driven BM-MI transitions are manifestations of the Lifshitz transition. Such feature seems to be general for HK-like models.\cite{CONTINENTINO1994619,PhysRevB.106.155119}

%Instead of a single point, there is a momentum region around $k_0$ that has the same and maximum distribution weight, suggesting localization in real space.

%As shown in fig.~\ref{fig:dos} (a-b), for $\mu/U=\frac{1}{2}$ with increasing $W/U$ the Boson density is invariant as $n=1$.
%Although Fig.~\ref{fig:dos} (a) and Fig.~\ref{fig:dos} (a) have the same number density, the band structure is different.
%For $\mu/U=\frac{1}{2}$ one fully filled band changes to two partially filled bands when crossing the critical $W_c/U=1$.
%This critical bandwidth will be further discussed in Sec.~\ref{sec3:level1}.

%Considering $q_x=0,q_y \rightarrow 0$,
%\begin{equation}
%	\begin{aligned}	
%		& \left[ j_x^p(\vec{q_y},\tau), j_x^p(\vec{-q_y},0)\right]=\\
%		&\frac{t^2}{N^2}\sum_{k_1,k_2}(e^{-i2k_x}+e^{i2k_x}-2)\left[ c_{k_1}^{\dag}(\tau)c_{k_1+q_y}(\tau),c_{k_2}^{\dag}(0)c_{k_2-q_y}(0)\right]
%		\label{eq:eq6}
%	\end{aligned}
%\end{equation}

%Here, $\Lambda_{xx}(\vec{q},i\omega_m)$ in the imaginary term is directly related to the susceptibility $\chi_{j_{x} j_{x}}(\vec{q},i\omega_m)$
% \begin{equation}
%	\begin{aligned}	
%		&\Lambda_{xx}(q_x=0,q_y \rightarrow 0, i\omega_m=0)=t{Re} \chi_{j_{x} j_{x}}(q_x=0,q_y \rightarrow 0, i\omega_m=0)\\
%		&\Lambda_{xx}(\vec{q=0},i\omega_m)=\text{Im} \chi_{j_{x} j_{x}}(\vec{q},\omega=0).
%		\label{eq:Chi}
%	\end{aligned}
%\end{equation}

\section{\label{sec4:level1}finite-temperature properties}

\begin{figure}
	\centering
	\includegraphics[width=0.68\columnwidth]{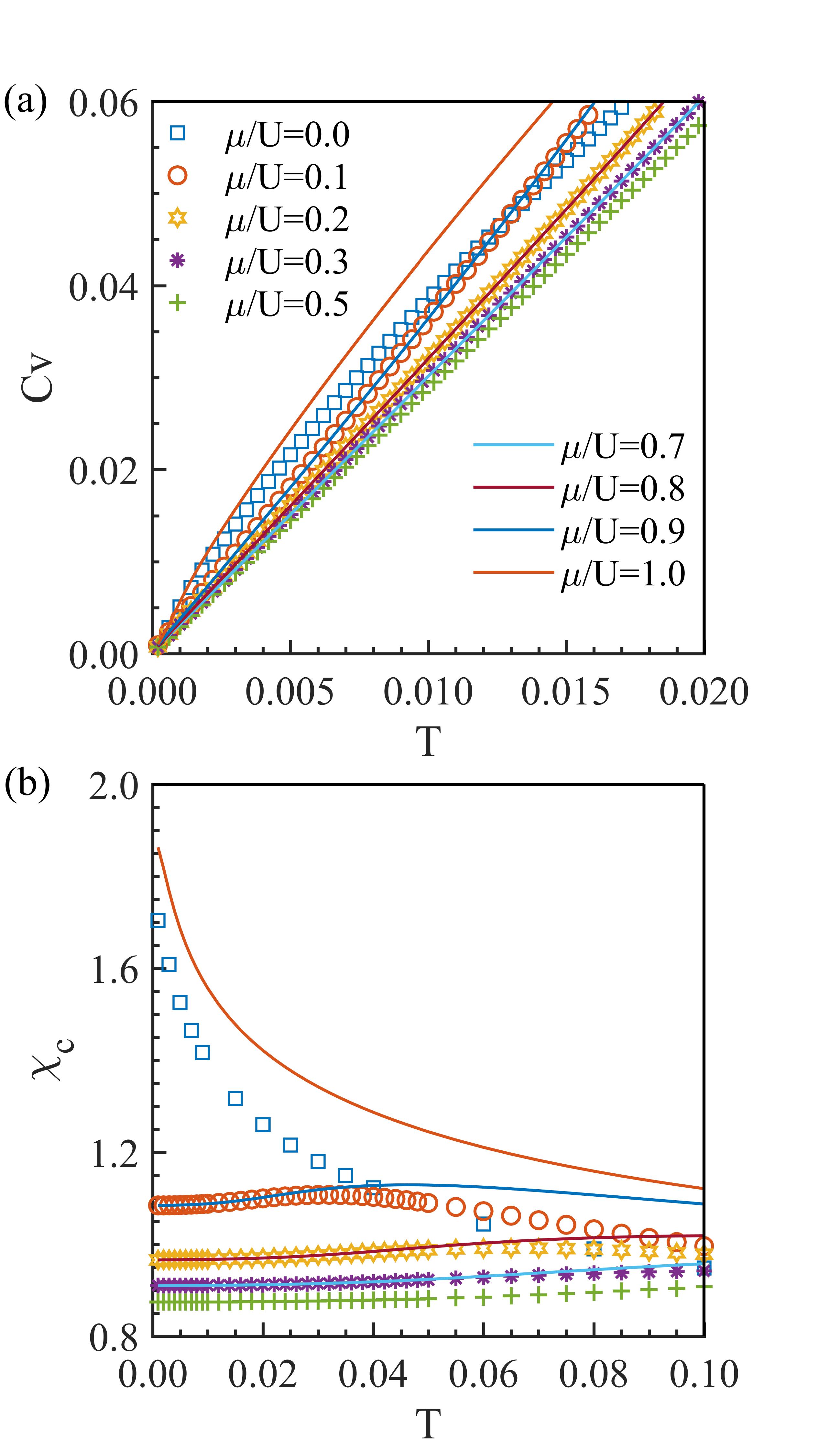}
	\caption{(a) The heat capacity $Cv$ and (b) the charge susceptibility $\chi_c$ versus temperature with various chemical potential $\mu$ at $W/U=2$ in the BM state.
	Except for the situation $\mu/U=0,1,...$, $Cv$ demonstrates a linear dependence when approaching zeros temperature, while $\chi_c$ is a constant number proportional to $N(0)$.}
	\label{fig:Cv}
\end{figure}

In this section, we investigate the heat capacity and the charge susceptibility at finite temperature.
Using the basis given in Eq.~\ref{eq:basis} and the energy spectrum, we can easily calculate the averages of observables using the partition function. For instance, the energy can be expressed as
\begin{equation}
	\begin{aligned}	
 E=\langle \hat{H} \rangle = \sum_{n}\frac{1}{\mathcal{Z}} e^{-\beta E_n} \langle n |\hat{H} |n\rangle.
		\label{eq:E}
	\end{aligned}
\end{equation}
As shown in Fig.~\ref{fig:Cv} (a), the heat capacity in the BM state exhibits a clear linear dependence on temperature, which is consistent with the behavior of a Fermi liquid ($Cv \sim \gamma T$).
A similar situation is observed for the charge susceptibility $\chi_c$.
In the BM states, $\chi_c$ saturates around zero temperature ($\chi_c(T\rightarrow 0) = c$). The constant $c$ shows a linear dependence on the particle density around the Bose surface ($c\sim N(0)$), as shown in Fig.~\ref{fig:Cv} (b). In a Fermi liquid, $N(0)$ represents the particle density around the Fermi surface.
Note that at $\mu/U=0,1$, $Cv$ deviates from the linear temperature dependence, and there is no saturation signal for $\chi_c$ at $T \rightarrow 0$ limit.
This deviation in the thermodynamic properties from the behavior of a Fermi liquid corresponds to the critical point of the Lifshitz transition between different BM states that we discussed earlier.

\section{\label{sec5:level1}Discussion}

\begin{figure}
	\centering
	\includegraphics[width=1.\columnwidth]{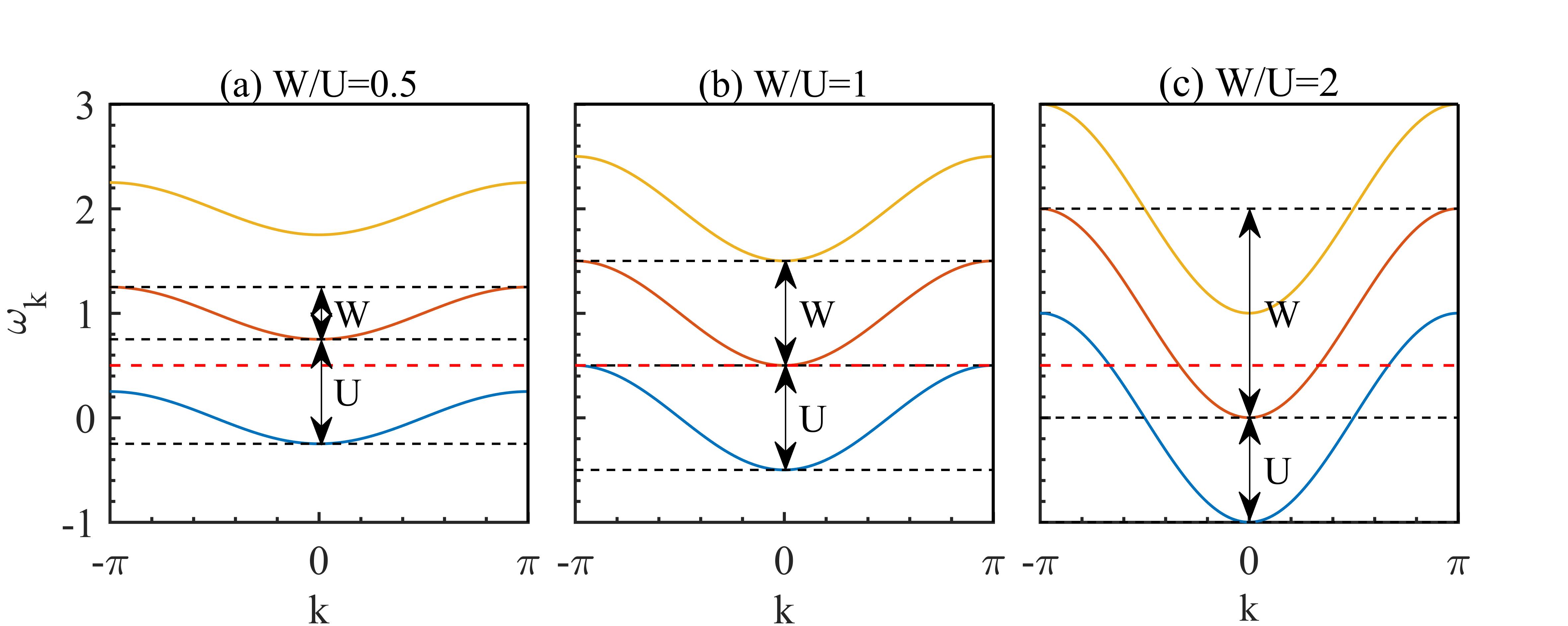}
	\caption{The schematic picture of the three lowest bands for (a) $W/U=0.5$; (b) the critical point $W_c/U=1$ (for the disappearing of MI state);
		(c) $W/U=2$. The chemical potential $\mu/U=0.5$ is denoted by red dashed line.}
	\label{fig:band}
\end{figure}

%\begin{figure}
%	\centering
%	\includegraphics[width=1.\columnwidth]{PSI2.jpg}
%	\caption{The ground-state energy versus field $\Psi$ (a) without and (b) with scattering term between different momentum in the BM state ($W/U=2, \mu/U=0.5$). }
%	\label{fig:PSI}
%\end{figure}

\subsection{Comparison with the Bose-Hubbard model}
In the BH model, a similar phase diagram is reported, where the MI state manifests in lobes with an integer boson density \cite{PhysRevB.70.024513,PhysRevB.61.12474,Sachdev}.
Here, we compare the properties of the MI and BM states in the BHK model with the MI and SF states in the BH model.

In both the BH and the BHK model, the MI state in the lobes is characterized by a finite energy gap to all excitations with no broken translation symmetry.
The total boson number $N$ is invariant under changes of chemical potential $\frac{\partial \langle \hat{N} \rangle}{\partial \mu}=0$,
demonstrating the incompressibility of the MI state.
Figure.~\ref{fig:band} depicts the schematic band structure of the BHK model.
The MI state exists only for $W/U<1$, where the chemical potential lies within the energy gap (see Fig.~\ref{fig:band}).
%Analogous to the BH model, the MI state in the BHK model can also be distinguished from the BM state by the incompressibility of total particle number, $\frac{\partial \langle \hat{N} \rangle}{\partial \mu}=0$.
The BHK system become compressible with the MI-BM transition. However, the robust direct gap existing in both MI and BM states yields the incompressibility of each independent $|{k} \rangle$ state, i.e., $\frac{\partial \langle \hat{n}_k \rangle}{\partial \mu}\neq 0$.
This precludes the SF, which relies on gapless excitations of the $\mid {k_0} \rangle$ state in the BH model.
Thus, we conclude that the BM is a metallic state with compressible wave-function and incompressible $\mid {k_0} \rangle$ state, distinguishing it from the SF state.
% The incompressiblility is intuitive in Fig.~\ref{fig:band}, where a direct energy gap at ${k_0}$ sustains in the BM state for any parameters.
%An extra information of Fig.~\ref{fig:band} is that $W_c/U=1$ is critical for any integer Boson density,
%which is the boundary of realizing the MI state.
%The previous work suggests a larger MI lobe for lower chemical potential \cite{CONTINENTINO1994619}.
%As shown in Fig.~\ref{fig:band}, only if $W=U$, the spectral weight is finite for any energy,
%the total Boson number will change under varying $\mu$.
%When increasing the bandwidth, the MI state can not sustain at $W/U>1$.

\subsection{Why BM state like Fermi liquid}
We have observed that the metallic states in BHK model, namely the BM state looks like the usual Fermi liquid because the former exhibits the linear-$T$ specific heat and nonzero $\chi_{c}$. To gain a more intuitive understanding of this phenomenon, let us consider the limit  $U\rightarrow\infty$ limit. In this scenario, the summation over $n_{k}$ in partition function $\mathcal{Z}=\prod_{k}\sum_{n_{k}=0}^{\infty}e^{-\beta E_{n_{k}}}$ is truncated to $n_{k}=1$. Consequently, we find $\mathcal{Z}=\prod_{k}(1+e^{-\beta (\varepsilon_{k}-\mu)})$ and its free energy is given by $F=-T\sum_{k}\ln(1+e^{-\beta (\varepsilon_{k}-\mu)})$.
This corresponds to a free fermion gas with dispersion $\varepsilon_{k}$. So, $Cv \sim T,\chi_{c}\sim$ constant
	has been explained.

Furthermore, in the $U\rightarrow\infty$ limit, the value of $n_{k}$ is either zero or one. For regimes where $n_{k}=0$, the Green's function only includes particle excitations, given by
	\begin{equation}
		\begin{aligned}	
			G^R(\omega,k)=\frac{1}{\omega-(\varepsilon_{k}-\mu)},
		\end{aligned}	
	\end{equation}
while for regimes where
$n_{k}=1$, only hole excitations exist, described by
\begin{equation}
		\begin{aligned}	
			G^R(\omega,k)=
			-\frac{1}{\omega-(\varepsilon_{k}-\mu)}.
		\end{aligned}	
	\end{equation}
It is interesting to note that the above Green's functions correspond to free fermion's counterpart, although the hole excitation in the BM state carries an additional minus sign. This sign is necessary to ensure causality in the retarded Green's function of any boson.

Finally, since the ground states of the BHK model are all product-states characterized by the occupation of each momentum $k$, and in the $U\rightarrow\infty$ limit, the state with $n_{k}=0$ ($n_{k}=1$) has energy $0$ ($\varepsilon_{k}-\mu$), there exist boundaries ($\varepsilon_{k}-\mu=0$) that separate unoccupied and occupied states. These boundaries serve as the Fermi surface and are the expected Bose surface in our model.

\section{\label{sec6:level1}conclusion}

In conclusion, our study has uncovered a MI-BM transition in the exactly solvable BHK model, which falls under the universality class of the Lifshitz transition.
The existence of the BM state is supported by the distinct momentum distribution function, the presence of a finite Drude weight, and the absence of SF weight.
At low temperatures, the BM state exhibits a linear-$T$ dependent heat capacity and a saturate charge susceptibility, demonstrating behavior akin to a Fermi liquid. Comparing the BM state with the SF state observed in the BH model,
we conclude that the BM state is characterized by a compressible total wave-function and an incompressible zero-momentum component.

Importantly, our paper presents a promising approach to realize this exotic BM state. It is indicated that any finite infinite-range interaction can disrupt the SF state, which aligns with the conventional notion that nontrivial long-range interaction is crucial for accessing the BM state. The contrasting nature between long-ranged and short-ranged interactions is clearly demonstrated in the different phase diagram in the BHK model with infinite-range interaction and the BH model with on-site interaction, leading to distinct stable phases with weak interaction, namely the BM state and the SF state, respectively. In some sense, the infinite-range HK interaction just frustrates bosons in momentum space, thus bosons are not likely to condensate into any particular momentum and no SF forms. 

However, we note that the BHK model studied here cannot give finite resistivity at finite temperature, which has been observed in experiments on boson metals, since no disorder or impurity effect is included. But, it is also noted that the existence of our BM states do nor require external magnetic field, disorder or fine-tuning of carrier density, thus BM states could be a robust state of matter, which is stabilized by nontrivial HK interaction. Therefore, we believe that the study of HK-like models is indeed a new direction in the many-body physics and many more interesting physics will be discovered when complicated details are included.

	\begin{acknowledgements}
We thank Z Yao for his useful comments.This research was supported in part by Supercomputing Center of
Lanzhou University and NSFC under Grant No.~$11834005$, No.~$11874188$.	
We thank the Supercomputing Center of Lanzhou University for allocation of CPU time.		
	\end{acknowledgements}

\bibliography{ref}

%merlin.mbs apsrev4-1.bst 2010-07-25 4.21a (PWD, AO, DPC) hacked
%Control: key (0)
%Control: author (72) initials jnrlst
%Control: editor formatted (1) identically to author
%Control: production of article title (-1) disabled
%Control: page (0) single
%Control: year (1) truncated
%Control: production of eprint (0) enabled
\begin{thebibliography}{76}%
\makeatletter
\providecommand \@ifxundefined [1]{%
 \@ifx{#1\undefined}
}%
\providecommand \@ifnum [1]{%
 \ifnum #1\expandafter \@firstoftwo
 \else \expandafter \@secondoftwo
 \fi
}%
\providecommand \@ifx [1]{%
 \ifx #1\expandafter \@firstoftwo
 \else \expandafter \@secondoftwo
 \fi
}%
\providecommand \natexlab [1]{#1}%
\providecommand \enquote  [1]{``#1''}%
\providecommand \bibnamefont  [1]{#1}%
\providecommand \bibfnamefont [1]{#1}%
\providecommand \citenamefont [1]{#1}%
\providecommand \href@noop [0]{\@secondoftwo}%
\providecommand \href [0]{\begingroup \@sanitize@url \@href}%
\providecommand \@href[1]{\@@startlink{#1}\@@href}%
\providecommand \@@href[1]{\endgroup#1\@@endlink}%
\providecommand \@sanitize@url [0]{\catcode `\\12\catcode `\$12\catcode
  `\&12\catcode `\#12\catcode `\^12\catcode `\_12\catcode `\%12\relax}%
\providecommand \@@startlink[1]{}%
\providecommand \@@endlink[0]{}%
\providecommand \url  [0]{\begingroup\@sanitize@url \@url }%
\providecommand \@url [1]{\endgroup\@href {#1}{\urlprefix }}%
\providecommand \urlprefix  [0]{URL }%
\providecommand \Eprint [0]{\href }%
\providecommand \doibase [0]{http://dx.doi.org/}%
\providecommand \selectlanguage [0]{\@gobble}%
\providecommand \bibinfo  [0]{\@secondoftwo}%
\providecommand \bibfield  [0]{\@secondoftwo}%
\providecommand \translation [1]{[#1]}%
\providecommand \BibitemOpen [0]{}%
\providecommand \bibitemStop [0]{}%
\providecommand \bibitemNoStop [0]{.\EOS\space}%
\providecommand \EOS [0]{\spacefactor3000\relax}%
\providecommand \BibitemShut  [1]{\csname bibitem#1\endcsname}%
\let\auto@bib@innerbib\@empty
%</preamble>
\bibitem [{\citenamefont {Fetter}\ and\ \citenamefont
  {Walecka}(2003)}]{fetter_quantum_1976}%
  \BibitemOpen
  \bibfield  {author} {\bibinfo {author} {\bibfnamefont {A.}~\bibnamefont
  {Fetter}}\ and\ \bibinfo {author} {\bibfnamefont {J.}~\bibnamefont
  {Walecka}},\ }\href {https://books.google.com.sg/books?id=0wekf1s83b0C}
  {\emph {\bibinfo {title} {Quantum Theory of Many-particle Systems}}},\ Dover
  Books on Physics\ (\bibinfo  {publisher} {Dover Publications},\ \bibinfo
  {year} {2003})\BibitemShut {NoStop}%
\bibitem [{\citenamefont {Greiner}\ \emph {et~al.}(2002)\citenamefont
  {Greiner}, \citenamefont {Mandel}, \citenamefont {Esslinger}, \citenamefont
  {H{\"a}nsch},\ and\ \citenamefont {Bloch}}]{greiner_quantum_2002}%
  \BibitemOpen
  \bibfield  {author} {\bibinfo {author} {\bibfnamefont {M.}~\bibnamefont
  {Greiner}}, \bibinfo {author} {\bibfnamefont {O.}~\bibnamefont {Mandel}},
  \bibinfo {author} {\bibfnamefont {T.}~\bibnamefont {Esslinger}}, \bibinfo
  {author} {\bibfnamefont {T.~W.}\ \bibnamefont {H{\"a}nsch}}, \ and\ \bibinfo
  {author} {\bibfnamefont {I.}~\bibnamefont {Bloch}},\ }\href {\doibase
  10.1038/415039a} {\bibfield  {journal} {\bibinfo  {journal} {Nature}\
  }\textbf {\bibinfo {volume} {415}},\ \bibinfo {pages} {39} (\bibinfo {year}
  {2002})}\BibitemShut {NoStop}%
\bibitem [{\citenamefont {Sachdev}(2023)}]{Sachdev}%
  \BibitemOpen
  \bibfield  {author} {\bibinfo {author} {\bibfnamefont {S.}~\bibnamefont
  {Sachdev}},\ }\href {\doibase 10.1017/9781009212717} {\emph {\bibinfo {title}
  {Quantum phases of matter}}}\ (\bibinfo  {publisher} {Cambridge University
  Press},\ \bibinfo {year} {2023})\BibitemShut {NoStop}%
\bibitem [{\citenamefont {Fisher}\ \emph {et~al.}(1989)\citenamefont {Fisher},
  \citenamefont {Weichman}, \citenamefont {Grinstein},\ and\ \citenamefont
  {Fisher}}]{fisher_boson_1989}%
  \BibitemOpen
  \bibfield  {author} {\bibinfo {author} {\bibfnamefont {M.~P.~A.}\
  \bibnamefont {Fisher}}, \bibinfo {author} {\bibfnamefont {P.~B.}\
  \bibnamefont {Weichman}}, \bibinfo {author} {\bibfnamefont {G.}~\bibnamefont
  {Grinstein}}, \ and\ \bibinfo {author} {\bibfnamefont {D.~S.}\ \bibnamefont
  {Fisher}},\ }\href {\doibase 10.1103/PhysRevB.40.546} {\bibfield  {journal}
  {\bibinfo  {journal} {Physical Review B}\ }\textbf {\bibinfo {volume} {40}},\
  \bibinfo {pages} {546} (\bibinfo {year} {1989})}\BibitemShut {NoStop}%
\bibitem [{\citenamefont {Abrahams}\ \emph {et~al.}(1979)\citenamefont
  {Abrahams}, \citenamefont {Anderson}, \citenamefont {Licciardello},\ and\
  \citenamefont {Ramakrishnan}}]{abrahams_scaling_1979}%
  \BibitemOpen
  \bibfield  {author} {\bibinfo {author} {\bibfnamefont {E.}~\bibnamefont
  {Abrahams}}, \bibinfo {author} {\bibfnamefont {P.~W.}\ \bibnamefont
  {Anderson}}, \bibinfo {author} {\bibfnamefont {D.~C.}\ \bibnamefont
  {Licciardello}}, \ and\ \bibinfo {author} {\bibfnamefont {T.~V.}\
  \bibnamefont {Ramakrishnan}},\ }\href {\doibase 10.1103/PhysRevLett.42.673}
  {\bibfield  {journal} {\bibinfo  {journal} {Physical Review Letters}\
  }\textbf {\bibinfo {volume} {42}},\ \bibinfo {pages} {673} (\bibinfo {year}
  {1979})}\BibitemShut {NoStop}%
\bibitem [{\citenamefont {Jaeger}\ \emph {et~al.}(1989)\citenamefont {Jaeger},
  \citenamefont {Haviland}, \citenamefont {Orr},\ and\ \citenamefont
  {Goldman}}]{PhysRevB.40.182}%
  \BibitemOpen
  \bibfield  {author} {\bibinfo {author} {\bibfnamefont {H.~M.}\ \bibnamefont
  {Jaeger}}, \bibinfo {author} {\bibfnamefont {D.~B.}\ \bibnamefont
  {Haviland}}, \bibinfo {author} {\bibfnamefont {B.~G.}\ \bibnamefont {Orr}}, \
  and\ \bibinfo {author} {\bibfnamefont {A.~M.}\ \bibnamefont {Goldman}},\
  }\href {\doibase 10.1103/PhysRevB.40.182} {\bibfield  {journal} {\bibinfo
  {journal} {Phys. Rev. B}\ }\textbf {\bibinfo {volume} {40}},\ \bibinfo
  {pages} {182} (\bibinfo {year} {1989})}\BibitemShut {NoStop}%
\bibitem [{\citenamefont {Ephron}\ \emph {et~al.}(1996)\citenamefont {Ephron},
  \citenamefont {Yazdani}, \citenamefont {Kapitulnik},\ and\ \citenamefont
  {Beasley}}]{PhysRevLett.76.1529}%
  \BibitemOpen
  \bibfield  {author} {\bibinfo {author} {\bibfnamefont {D.}~\bibnamefont
  {Ephron}}, \bibinfo {author} {\bibfnamefont {A.}~\bibnamefont {Yazdani}},
  \bibinfo {author} {\bibfnamefont {A.}~\bibnamefont {Kapitulnik}}, \ and\
  \bibinfo {author} {\bibfnamefont {M.~R.}\ \bibnamefont {Beasley}},\ }\href
  {\doibase 10.1103/PhysRevLett.76.1529} {\bibfield  {journal} {\bibinfo
  {journal} {Phys. Rev. Lett.}\ }\textbf {\bibinfo {volume} {76}},\ \bibinfo
  {pages} {1529} (\bibinfo {year} {1996})}\BibitemShut {NoStop}%
\bibitem [{\citenamefont {Mason}\ and\ \citenamefont
  {Kapitulnik}(1999)}]{PhysRevLett.82.5341}%
  \BibitemOpen
  \bibfield  {author} {\bibinfo {author} {\bibfnamefont {N.}~\bibnamefont
  {Mason}}\ and\ \bibinfo {author} {\bibfnamefont {A.}~\bibnamefont
  {Kapitulnik}},\ }\href {\doibase 10.1103/PhysRevLett.82.5341} {\bibfield
  {journal} {\bibinfo  {journal} {Phys. Rev. Lett.}\ }\textbf {\bibinfo
  {volume} {82}},\ \bibinfo {pages} {5341} (\bibinfo {year}
  {1999})}\BibitemShut {NoStop}%
\bibitem [{\citenamefont {Christiansen}\ \emph {et~al.}(2002)\citenamefont
  {Christiansen}, \citenamefont {Hernandez},\ and\ \citenamefont
  {Goldman}}]{PhysRevLett.88.037004}%
  \BibitemOpen
  \bibfield  {author} {\bibinfo {author} {\bibfnamefont {C.}~\bibnamefont
  {Christiansen}}, \bibinfo {author} {\bibfnamefont {L.~M.}\ \bibnamefont
  {Hernandez}}, \ and\ \bibinfo {author} {\bibfnamefont {A.~M.}\ \bibnamefont
  {Goldman}},\ }\href {\doibase 10.1103/PhysRevLett.88.037004} {\bibfield
  {journal} {\bibinfo  {journal} {Phys. Rev. Lett.}\ }\textbf {\bibinfo
  {volume} {88}},\ \bibinfo {pages} {037004} (\bibinfo {year}
  {2002})}\BibitemShut {NoStop}%
\bibitem [{\citenamefont {Eley}\ \emph {et~al.}(2012)\citenamefont {Eley},
  \citenamefont {Gopalakrishnan}, \citenamefont {Goldbart},\ and\ \citenamefont
  {Mason}}]{Eley2012}%
  \BibitemOpen
  \bibfield  {author} {\bibinfo {author} {\bibfnamefont {S.}~\bibnamefont
  {Eley}}, \bibinfo {author} {\bibfnamefont {S.}~\bibnamefont
  {Gopalakrishnan}}, \bibinfo {author} {\bibfnamefont {P.~M.}\ \bibnamefont
  {Goldbart}}, \ and\ \bibinfo {author} {\bibfnamefont {N.}~\bibnamefont
  {Mason}},\ }\href {\doibase 10.1038/nphys2154} {\bibfield  {journal}
  {\bibinfo  {journal} {Nature Physics}\ }\textbf {\bibinfo {volume} {1}},\
  \bibinfo {pages} {59} (\bibinfo {year} {2012})}\BibitemShut {NoStop}%
\bibitem [{\citenamefont {Liu}\ \emph {et~al.}(2013)\citenamefont {Liu},
  \citenamefont {Pan}, \citenamefont {Wen}, \citenamefont {Kim}, \citenamefont
  {Sambandamurthy},\ and\ \citenamefont {Armitage}}]{PhysRevLett.111.067003}%
  \BibitemOpen
  \bibfield  {author} {\bibinfo {author} {\bibfnamefont {W.}~\bibnamefont
  {Liu}}, \bibinfo {author} {\bibfnamefont {L.}~\bibnamefont {Pan}}, \bibinfo
  {author} {\bibfnamefont {J.}~\bibnamefont {Wen}}, \bibinfo {author}
  {\bibfnamefont {M.}~\bibnamefont {Kim}}, \bibinfo {author} {\bibfnamefont
  {G.}~\bibnamefont {Sambandamurthy}}, \ and\ \bibinfo {author} {\bibfnamefont
  {N.~P.}\ \bibnamefont {Armitage}},\ }\href {\doibase
  10.1103/PhysRevLett.111.067003} {\bibfield  {journal} {\bibinfo  {journal}
  {Phys. Rev. Lett.}\ }\textbf {\bibinfo {volume} {111}},\ \bibinfo {pages}
  {067003} (\bibinfo {year} {2013})}\BibitemShut {NoStop}%
\bibitem [{\citenamefont {Garcia-Barriocanal}\ \emph
  {et~al.}(2013)\citenamefont {Garcia-Barriocanal}, \citenamefont {Kobrinskii},
  \citenamefont {Leng}, \citenamefont {Kinney}, \citenamefont {Yang},
  \citenamefont {Snyder},\ and\ \citenamefont {Goldman}}]{PhysRevB.87.024509}%
  \BibitemOpen
  \bibfield  {author} {\bibinfo {author} {\bibfnamefont {J.}~\bibnamefont
  {Garcia-Barriocanal}}, \bibinfo {author} {\bibfnamefont {A.}~\bibnamefont
  {Kobrinskii}}, \bibinfo {author} {\bibfnamefont {X.}~\bibnamefont {Leng}},
  \bibinfo {author} {\bibfnamefont {J.}~\bibnamefont {Kinney}}, \bibinfo
  {author} {\bibfnamefont {B.}~\bibnamefont {Yang}}, \bibinfo {author}
  {\bibfnamefont {S.}~\bibnamefont {Snyder}}, \ and\ \bibinfo {author}
  {\bibfnamefont {A.~M.}\ \bibnamefont {Goldman}},\ }\href {\doibase
  10.1103/PhysRevB.87.024509} {\bibfield  {journal} {\bibinfo  {journal} {Phys.
  Rev. B}\ }\textbf {\bibinfo {volume} {87}},\ \bibinfo {pages} {024509}
  (\bibinfo {year} {2013})}\BibitemShut {NoStop}%
\bibitem [{\citenamefont {Han}\ \emph {et~al.}(2014)\citenamefont {Han},
  \citenamefont {Allain}, \citenamefont {Arjmandi-Tash}, \citenamefont
  {Tikhonov}, \citenamefont {Feigel’man}, \citenamefont {Sacépé},\ and\
  \citenamefont {Bouchiat}}]{Han2014}%
  \BibitemOpen
  \bibfield  {author} {\bibinfo {author} {\bibfnamefont {Z.}~\bibnamefont
  {Han}}, \bibinfo {author} {\bibfnamefont {A.}~\bibnamefont {Allain}},
  \bibinfo {author} {\bibfnamefont {H.}~\bibnamefont {Arjmandi-Tash}}, \bibinfo
  {author} {\bibfnamefont {K.}~\bibnamefont {Tikhonov}}, \bibinfo {author}
  {\bibfnamefont {M.}~\bibnamefont {Feigel’man}}, \bibinfo {author}
  {\bibfnamefont {B.}~\bibnamefont {Sacépé}}, \ and\ \bibinfo {author}
  {\bibfnamefont {V.}~\bibnamefont {Bouchiat}},\ }\href {\doibase
  10.1038/nphys2929} {\bibfield  {journal} {\bibinfo  {journal} {Nature
  Physics}\ }\textbf {\bibinfo {volume} {10}},\ \bibinfo {pages} {380}
  (\bibinfo {year} {2014})}\BibitemShut {NoStop}%
\bibitem [{\citenamefont {Saito}\ \emph {et~al.}(2015)\citenamefont {Saito},
  \citenamefont {Kasahara}, \citenamefont {Ye}, \citenamefont {Iwasa},\ and\
  \citenamefont {Nojima}}]{doi:10.1126/science.1259440}%
  \BibitemOpen
  \bibfield  {author} {\bibinfo {author} {\bibfnamefont {Y.}~\bibnamefont
  {Saito}}, \bibinfo {author} {\bibfnamefont {Y.}~\bibnamefont {Kasahara}},
  \bibinfo {author} {\bibfnamefont {J.}~\bibnamefont {Ye}}, \bibinfo {author}
  {\bibfnamefont {Y.}~\bibnamefont {Iwasa}}, \ and\ \bibinfo {author}
  {\bibfnamefont {T.}~\bibnamefont {Nojima}},\ }\href {\doibase
  10.1126/science.1259440} {\bibfield  {journal} {\bibinfo  {journal}
  {Science}\ }\textbf {\bibinfo {volume} {350}},\ \bibinfo {pages} {409}
  (\bibinfo {year} {2015})},\ \Eprint
  {http://arxiv.org/abs/https://www.science.org/doi/pdf/10.1126/science.1259440}
  {https://www.science.org/doi/pdf/10.1126/science.1259440} \BibitemShut
  {NoStop}%
\bibitem [{\citenamefont {Breznay}\ and\ \citenamefont
  {Kapitulnik}(2017)}]{doi:10.1126/sciadv.1700612}%
  \BibitemOpen
  \bibfield  {author} {\bibinfo {author} {\bibfnamefont {N.~P.}\ \bibnamefont
  {Breznay}}\ and\ \bibinfo {author} {\bibfnamefont {A.}~\bibnamefont
  {Kapitulnik}},\ }\href {\doibase 10.1126/sciadv.1700612} {\bibfield
  {journal} {\bibinfo  {journal} {Science Advances}\ }\textbf {\bibinfo
  {volume} {3}},\ \bibinfo {pages} {e1700612} (\bibinfo {year} {2017})},\
  \Eprint
  {http://arxiv.org/abs/https://www.science.org/doi/pdf/10.1126/sciadv.1700612}
  {https://www.science.org/doi/pdf/10.1126/sciadv.1700612} \BibitemShut
  {NoStop}%
\bibitem [{\citenamefont {Kapitulnik}\ \emph {et~al.}(2019)\citenamefont
  {Kapitulnik}, \citenamefont {Kivelson},\ and\ \citenamefont
  {Spivak}}]{RevModPhys.91.011002}%
  \BibitemOpen
  \bibfield  {author} {\bibinfo {author} {\bibfnamefont {A.}~\bibnamefont
  {Kapitulnik}}, \bibinfo {author} {\bibfnamefont {S.~A.}\ \bibnamefont
  {Kivelson}}, \ and\ \bibinfo {author} {\bibfnamefont {B.}~\bibnamefont
  {Spivak}},\ }\href {\doibase 10.1103/RevModPhys.91.011002} {\bibfield
  {journal} {\bibinfo  {journal} {Rev. Mod. Phys.}\ }\textbf {\bibinfo {volume}
  {91}},\ \bibinfo {pages} {011002} (\bibinfo {year} {2019})}\BibitemShut
  {NoStop}%
\bibitem [{\citenamefont {Bøttcher}\ \emph {et~al.}(2018)\citenamefont
  {Bøttcher}, \citenamefont {Nichele}, \citenamefont {Kjaergaard},
  \citenamefont {Suominen}, \citenamefont {Shabani}, \citenamefont
  {Palmstrøm},\ and\ \citenamefont {Marcus}}]{Bottcher2018}%
  \BibitemOpen
  \bibfield  {author} {\bibinfo {author} {\bibfnamefont {C.~G.~L.}\
  \bibnamefont {Bøttcher}}, \bibinfo {author} {\bibfnamefont {F.}~\bibnamefont
  {Nichele}}, \bibinfo {author} {\bibfnamefont {M.}~\bibnamefont {Kjaergaard}},
  \bibinfo {author} {\bibfnamefont {H.~J.}\ \bibnamefont {Suominen}}, \bibinfo
  {author} {\bibfnamefont {J.}~\bibnamefont {Shabani}}, \bibinfo {author}
  {\bibfnamefont {C.~J.}\ \bibnamefont {Palmstrøm}}, \ and\ \bibinfo {author}
  {\bibfnamefont {C.~M.}\ \bibnamefont {Marcus}},\ }\href {\doibase
  10.1038/s41567-018-0259-9} {\bibfield  {journal} {\bibinfo  {journal} {Nature
  Physics}\ }\textbf {\bibinfo {volume} {14}},\ \bibinfo {pages} {1138}
  (\bibinfo {year} {2018})}\BibitemShut {NoStop}%
\bibitem [{\citenamefont {Chen}\ \emph
  {et~al.}(2018{\natexlab{a}})\citenamefont {Chen}, \citenamefont {Swartz},
  \citenamefont {Yoon}, \citenamefont {Inoue}, \citenamefont {Merz},
  \citenamefont {Lu}, \citenamefont {Xie}, \citenamefont {Yuan}, \citenamefont
  {Hikita}, \citenamefont {Raghu},\ and\ \citenamefont {Hwang}}]{Chen2018}%
  \BibitemOpen
  \bibfield  {author} {\bibinfo {author} {\bibfnamefont {Z.}~\bibnamefont
  {Chen}}, \bibinfo {author} {\bibfnamefont {A.~G.}\ \bibnamefont {Swartz}},
  \bibinfo {author} {\bibfnamefont {H.}~\bibnamefont {Yoon}}, \bibinfo {author}
  {\bibfnamefont {H.}~\bibnamefont {Inoue}}, \bibinfo {author} {\bibfnamefont
  {T.~A.}\ \bibnamefont {Merz}}, \bibinfo {author} {\bibfnamefont
  {D.}~\bibnamefont {Lu}}, \bibinfo {author} {\bibfnamefont {Y.}~\bibnamefont
  {Xie}}, \bibinfo {author} {\bibfnamefont {H.}~\bibnamefont {Yuan}}, \bibinfo
  {author} {\bibfnamefont {Y.}~\bibnamefont {Hikita}}, \bibinfo {author}
  {\bibfnamefont {S.}~\bibnamefont {Raghu}}, \ and\ \bibinfo {author}
  {\bibfnamefont {H.~Y.}\ \bibnamefont {Hwang}},\ }\href {\doibase
  10.1038/s41467-018-06444-2} {\bibfield  {journal} {\bibinfo  {journal}
  {Nature Communications}\ }\textbf {\bibinfo {volume} {9}},\ \bibinfo {pages}
  {4008} (\bibinfo {year} {2018}{\natexlab{a}})}\BibitemShut {NoStop}%
\bibitem [{\citenamefont {Yang}\ \emph {et~al.}(2019)\citenamefont {Yang},
  \citenamefont {Liu}, \citenamefont {Wang}, \citenamefont {Feng},
  \citenamefont {He}, \citenamefont {Sun}, \citenamefont {Tang}, \citenamefont
  {Wu}, \citenamefont {Xiong}, \citenamefont {Zhang}, \citenamefont {Lin},
  \citenamefont {Yao}, \citenamefont {Liu}, \citenamefont {Fernandes},
  \citenamefont {Xu}, \citenamefont {Valles}, \citenamefont {Wang},\ and\
  \citenamefont {Li}}]{doi:10.1126/science.aax5798}%
  \BibitemOpen
  \bibfield  {author} {\bibinfo {author} {\bibfnamefont {C.}~\bibnamefont
  {Yang}}, \bibinfo {author} {\bibfnamefont {Y.}~\bibnamefont {Liu}}, \bibinfo
  {author} {\bibfnamefont {Y.}~\bibnamefont {Wang}}, \bibinfo {author}
  {\bibfnamefont {L.}~\bibnamefont {Feng}}, \bibinfo {author} {\bibfnamefont
  {Q.}~\bibnamefont {He}}, \bibinfo {author} {\bibfnamefont {J.}~\bibnamefont
  {Sun}}, \bibinfo {author} {\bibfnamefont {Y.}~\bibnamefont {Tang}}, \bibinfo
  {author} {\bibfnamefont {C.}~\bibnamefont {Wu}}, \bibinfo {author}
  {\bibfnamefont {J.}~\bibnamefont {Xiong}}, \bibinfo {author} {\bibfnamefont
  {W.}~\bibnamefont {Zhang}}, \bibinfo {author} {\bibfnamefont
  {X.}~\bibnamefont {Lin}}, \bibinfo {author} {\bibfnamefont {H.}~\bibnamefont
  {Yao}}, \bibinfo {author} {\bibfnamefont {H.}~\bibnamefont {Liu}}, \bibinfo
  {author} {\bibfnamefont {G.}~\bibnamefont {Fernandes}}, \bibinfo {author}
  {\bibfnamefont {J.}~\bibnamefont {Xu}}, \bibinfo {author} {\bibfnamefont
  {J.~M.}\ \bibnamefont {Valles}}, \bibinfo {author} {\bibfnamefont
  {J.}~\bibnamefont {Wang}}, \ and\ \bibinfo {author} {\bibfnamefont
  {Y.}~\bibnamefont {Li}},\ }\href {\doibase 10.1126/science.aax5798}
  {\bibfield  {journal} {\bibinfo  {journal} {Science}\ }\textbf {\bibinfo
  {volume} {366}},\ \bibinfo {pages} {1505} (\bibinfo {year} {2019})},\ \Eprint
  {http://arxiv.org/abs/https://www.science.org/doi/pdf/10.1126/science.aax5798}
  {https://www.science.org/doi/pdf/10.1126/science.aax5798} \BibitemShut
  {NoStop}%
\bibitem [{\citenamefont {Tsen}\ \emph {et~al.}(2016)\citenamefont {Tsen},
  \citenamefont {Hunt}, \citenamefont {Kim}, \citenamefont {Yuan},
  \citenamefont {Jia}, \citenamefont {Cava}, \citenamefont {Hone},
  \citenamefont {Kim}, \citenamefont {Dean},\ and\ \citenamefont
  {Pasupathy}}]{Tsen2016}%
  \BibitemOpen
  \bibfield  {author} {\bibinfo {author} {\bibfnamefont {A.~W.}\ \bibnamefont
  {Tsen}}, \bibinfo {author} {\bibfnamefont {B.}~\bibnamefont {Hunt}}, \bibinfo
  {author} {\bibfnamefont {Y.~D.}\ \bibnamefont {Kim}}, \bibinfo {author}
  {\bibfnamefont {Z.~J.}\ \bibnamefont {Yuan}}, \bibinfo {author}
  {\bibfnamefont {S.}~\bibnamefont {Jia}}, \bibinfo {author} {\bibfnamefont
  {R.~J.}\ \bibnamefont {Cava}}, \bibinfo {author} {\bibfnamefont
  {J.}~\bibnamefont {Hone}}, \bibinfo {author} {\bibfnamefont {P.}~\bibnamefont
  {Kim}}, \bibinfo {author} {\bibfnamefont {C.~R.}\ \bibnamefont {Dean}}, \
  and\ \bibinfo {author} {\bibfnamefont {A.~N.}\ \bibnamefont {Pasupathy}},\
  }\href {\doibase 10.1038/nphys3579} {\bibfield  {journal} {\bibinfo
  {journal} {Nature Physics}\ }\textbf {\bibinfo {volume} {12}},\ \bibinfo
  {pages} {208} (\bibinfo {year} {2016})}\BibitemShut {NoStop}%
\bibitem [{\citenamefont {Shimshoni}\ \emph {et~al.}(1998)\citenamefont
  {Shimshoni}, \citenamefont {Auerbach},\ and\ \citenamefont
  {Kapitulnik}}]{PhysRevLett.80.3352}%
  \BibitemOpen
  \bibfield  {author} {\bibinfo {author} {\bibfnamefont {E.}~\bibnamefont
  {Shimshoni}}, \bibinfo {author} {\bibfnamefont {A.}~\bibnamefont {Auerbach}},
  \ and\ \bibinfo {author} {\bibfnamefont {A.}~\bibnamefont {Kapitulnik}},\
  }\href {\doibase 10.1103/PhysRevLett.80.3352} {\bibfield  {journal} {\bibinfo
   {journal} {Phys. Rev. Lett.}\ }\textbf {\bibinfo {volume} {80}},\ \bibinfo
  {pages} {3352} (\bibinfo {year} {1998})}\BibitemShut {NoStop}%
\bibitem [{\citenamefont {Kapitulnik}\ \emph {et~al.}(2001)\citenamefont
  {Kapitulnik}, \citenamefont {Mason}, \citenamefont {Kivelson},\ and\
  \citenamefont {Chakravarty}}]{PhysRevB.63.125322}%
  \BibitemOpen
  \bibfield  {author} {\bibinfo {author} {\bibfnamefont {A.}~\bibnamefont
  {Kapitulnik}}, \bibinfo {author} {\bibfnamefont {N.}~\bibnamefont {Mason}},
  \bibinfo {author} {\bibfnamefont {S.~A.}\ \bibnamefont {Kivelson}}, \ and\
  \bibinfo {author} {\bibfnamefont {S.}~\bibnamefont {Chakravarty}},\ }\href
  {\doibase 10.1103/PhysRevB.63.125322} {\bibfield  {journal} {\bibinfo
  {journal} {Phys. Rev. B}\ }\textbf {\bibinfo {volume} {63}},\ \bibinfo
  {pages} {125322} (\bibinfo {year} {2001})}\BibitemShut {NoStop}%
\bibitem [{\citenamefont {Galitski}\ \emph {et~al.}(2005)\citenamefont
  {Galitski}, \citenamefont {Refael}, \citenamefont {Fisher},\ and\
  \citenamefont {Senthil}}]{PhysRevLett.95.077002}%
  \BibitemOpen
  \bibfield  {author} {\bibinfo {author} {\bibfnamefont {V.~M.}\ \bibnamefont
  {Galitski}}, \bibinfo {author} {\bibfnamefont {G.}~\bibnamefont {Refael}},
  \bibinfo {author} {\bibfnamefont {M.~P.~A.}\ \bibnamefont {Fisher}}, \ and\
  \bibinfo {author} {\bibfnamefont {T.}~\bibnamefont {Senthil}},\ }\href
  {\doibase 10.1103/PhysRevLett.95.077002} {\bibfield  {journal} {\bibinfo
  {journal} {Phys. Rev. Lett.}\ }\textbf {\bibinfo {volume} {95}},\ \bibinfo
  {pages} {077002} (\bibinfo {year} {2005})}\BibitemShut {NoStop}%
\bibitem [{\citenamefont {Lee}\ \emph {et~al.}(1991)\citenamefont {Lee},
  \citenamefont {Kivelson},\ and\ \citenamefont {Zhang}}]{PhysRevLett.67.3302}%
  \BibitemOpen
  \bibfield  {author} {\bibinfo {author} {\bibfnamefont {D.-H.}\ \bibnamefont
  {Lee}}, \bibinfo {author} {\bibfnamefont {S.}~\bibnamefont {Kivelson}}, \
  and\ \bibinfo {author} {\bibfnamefont {S.-C.}\ \bibnamefont {Zhang}},\ }\href
  {\doibase 10.1103/PhysRevLett.67.3302} {\bibfield  {journal} {\bibinfo
  {journal} {Phys. Rev. Lett.}\ }\textbf {\bibinfo {volume} {67}},\ \bibinfo
  {pages} {3302} (\bibinfo {year} {1991})}\BibitemShut {NoStop}%
\bibitem [{\citenamefont {Das}\ and\ \citenamefont
  {Doniach}(1999)}]{PhysRevB.60.1261}%
  \BibitemOpen
  \bibfield  {author} {\bibinfo {author} {\bibfnamefont {D.}~\bibnamefont
  {Das}}\ and\ \bibinfo {author} {\bibfnamefont {S.}~\bibnamefont {Doniach}},\
  }\href {\doibase 10.1103/PhysRevB.60.1261} {\bibfield  {journal} {\bibinfo
  {journal} {Phys. Rev. B}\ }\textbf {\bibinfo {volume} {60}},\ \bibinfo
  {pages} {1261} (\bibinfo {year} {1999})}\BibitemShut {NoStop}%
\bibitem [{\citenamefont {Phillips}\ and\ \citenamefont
  {Dalidovich}(2003)}]{doi:10.1126/science.1088253}%
  \BibitemOpen
  \bibfield  {author} {\bibinfo {author} {\bibfnamefont {P.}~\bibnamefont
  {Phillips}}\ and\ \bibinfo {author} {\bibfnamefont {D.}~\bibnamefont
  {Dalidovich}},\ }\href {\doibase 10.1126/science.1088253} {\bibfield
  {journal} {\bibinfo  {journal} {Science}\ }\textbf {\bibinfo {volume}
  {302}},\ \bibinfo {pages} {243} (\bibinfo {year} {2003})},\ \Eprint
  {http://arxiv.org/abs/https://www.science.org/doi/pdf/10.1126/science.1088253}
  {https://www.science.org/doi/pdf/10.1126/science.1088253} \BibitemShut
  {NoStop}%
\bibitem [{\citenamefont {Spivak}\ \emph {et~al.}(2001)\citenamefont {Spivak},
  \citenamefont {Zyuzin},\ and\ \citenamefont {Hruska}}]{PhysRevB.64.132502}%
  \BibitemOpen
  \bibfield  {author} {\bibinfo {author} {\bibfnamefont {B.}~\bibnamefont
  {Spivak}}, \bibinfo {author} {\bibfnamefont {A.}~\bibnamefont {Zyuzin}}, \
  and\ \bibinfo {author} {\bibfnamefont {M.}~\bibnamefont {Hruska}},\ }\href
  {\doibase 10.1103/PhysRevB.64.132502} {\bibfield  {journal} {\bibinfo
  {journal} {Phys. Rev. B}\ }\textbf {\bibinfo {volume} {64}},\ \bibinfo
  {pages} {132502} (\bibinfo {year} {2001})}\BibitemShut {NoStop}%
\bibitem [{\citenamefont {Spivak}\ \emph {et~al.}(2008)\citenamefont {Spivak},
  \citenamefont {Oreto},\ and\ \citenamefont {Kivelson}}]{PhysRevB.77.214523}%
  \BibitemOpen
  \bibfield  {author} {\bibinfo {author} {\bibfnamefont {B.}~\bibnamefont
  {Spivak}}, \bibinfo {author} {\bibfnamefont {P.}~\bibnamefont {Oreto}}, \
  and\ \bibinfo {author} {\bibfnamefont {S.~A.}\ \bibnamefont {Kivelson}},\
  }\href {\doibase 10.1103/PhysRevB.77.214523} {\bibfield  {journal} {\bibinfo
  {journal} {Phys. Rev. B}\ }\textbf {\bibinfo {volume} {77}},\ \bibinfo
  {pages} {214523} (\bibinfo {year} {2008})}\BibitemShut {NoStop}%
\bibitem [{\citenamefont {Mulligan}\ and\ \citenamefont
  {Raghu}(2016)}]{PhysRevB.93.205116}%
  \BibitemOpen
  \bibfield  {author} {\bibinfo {author} {\bibfnamefont {M.}~\bibnamefont
  {Mulligan}}\ and\ \bibinfo {author} {\bibfnamefont {S.}~\bibnamefont
  {Raghu}},\ }\href {\doibase 10.1103/PhysRevB.93.205116} {\bibfield  {journal}
  {\bibinfo  {journal} {Phys. Rev. B}\ }\textbf {\bibinfo {volume} {93}},\
  \bibinfo {pages} {205116} (\bibinfo {year} {2016})}\BibitemShut {NoStop}%
\bibitem [{\citenamefont {Feigel'man}\ and\ \citenamefont
  {Larkin}(1998)}]{FEIGELMAN1998107}%
  \BibitemOpen
  \bibfield  {author} {\bibinfo {author} {\bibfnamefont {M.}~\bibnamefont
  {Feigel'man}}\ and\ \bibinfo {author} {\bibfnamefont {A.}~\bibnamefont
  {Larkin}},\ }\href {\doibase https://doi.org/10.1016/S0301-0104(98)00075-5}
  {\bibfield  {journal} {\bibinfo  {journal} {Chemical Physics}\ }\textbf
  {\bibinfo {volume} {235}},\ \bibinfo {pages} {107} (\bibinfo {year}
  {1998})}\BibitemShut {NoStop}%
\bibitem [{\citenamefont {Dalidovich}\ and\ \citenamefont
  {Phillips}(2002)}]{PhysRevLett.89.027001}%
  \BibitemOpen
  \bibfield  {author} {\bibinfo {author} {\bibfnamefont {D.}~\bibnamefont
  {Dalidovich}}\ and\ \bibinfo {author} {\bibfnamefont {P.}~\bibnamefont
  {Phillips}},\ }\href {\doibase 10.1103/PhysRevLett.89.027001} {\bibfield
  {journal} {\bibinfo  {journal} {Phys. Rev. Lett.}\ }\textbf {\bibinfo
  {volume} {89}},\ \bibinfo {pages} {027001} (\bibinfo {year}
  {2002})}\BibitemShut {NoStop}%
\bibitem [{\citenamefont {Wu}\ and\ \citenamefont
  {Phillips}(2006)}]{PhysRevB.73.214507}%
  \BibitemOpen
  \bibfield  {author} {\bibinfo {author} {\bibfnamefont {J.}~\bibnamefont
  {Wu}}\ and\ \bibinfo {author} {\bibfnamefont {P.}~\bibnamefont {Phillips}},\
  }\href {\doibase 10.1103/PhysRevB.73.214507} {\bibfield  {journal} {\bibinfo
  {journal} {Phys. Rev. B}\ }\textbf {\bibinfo {volume} {73}},\ \bibinfo
  {pages} {214507} (\bibinfo {year} {2006})}\BibitemShut {NoStop}%
\bibitem [{\citenamefont {Davison}\ \emph {et~al.}(2016)\citenamefont
  {Davison}, \citenamefont {Delacr\'etaz}, \citenamefont {Gout\'eraux},\ and\
  \citenamefont {Hartnoll}}]{PhysRevB.94.054502}%
  \BibitemOpen
  \bibfield  {author} {\bibinfo {author} {\bibfnamefont {R.~A.}\ \bibnamefont
  {Davison}}, \bibinfo {author} {\bibfnamefont {L.~V.}\ \bibnamefont
  {Delacr\'etaz}}, \bibinfo {author} {\bibfnamefont {B.}~\bibnamefont
  {Gout\'eraux}}, \ and\ \bibinfo {author} {\bibfnamefont {S.~A.}\ \bibnamefont
  {Hartnoll}},\ }\href {\doibase 10.1103/PhysRevB.94.054502} {\bibfield
  {journal} {\bibinfo  {journal} {Phys. Rev. B}\ }\textbf {\bibinfo {volume}
  {94}},\ \bibinfo {pages} {054502} (\bibinfo {year} {2016})}\BibitemShut
  {NoStop}%
\bibitem [{\citenamefont {Das}\ and\ \citenamefont
  {Doniach}(2001)}]{PhysRevB.64.134511}%
  \BibitemOpen
  \bibfield  {author} {\bibinfo {author} {\bibfnamefont {D.}~\bibnamefont
  {Das}}\ and\ \bibinfo {author} {\bibfnamefont {S.}~\bibnamefont {Doniach}},\
  }\href {\doibase 10.1103/PhysRevB.64.134511} {\bibfield  {journal} {\bibinfo
  {journal} {Phys. Rev. B}\ }\textbf {\bibinfo {volume} {64}},\ \bibinfo
  {pages} {134511} (\bibinfo {year} {2001})}\BibitemShut {NoStop}%
\bibitem [{\citenamefont {van~der Zant}\ \emph {et~al.}(1996)\citenamefont
  {van~der Zant}, \citenamefont {Elion}, \citenamefont {Geerligs},\ and\
  \citenamefont {Mooij}}]{PhysRevB.54.10081}%
  \BibitemOpen
  \bibfield  {author} {\bibinfo {author} {\bibfnamefont {H.~S.~J.}\
  \bibnamefont {van~der Zant}}, \bibinfo {author} {\bibfnamefont {W.~J.}\
  \bibnamefont {Elion}}, \bibinfo {author} {\bibfnamefont {L.~J.}\ \bibnamefont
  {Geerligs}}, \ and\ \bibinfo {author} {\bibfnamefont {J.~E.}\ \bibnamefont
  {Mooij}},\ }\href {\doibase 10.1103/PhysRevB.54.10081} {\bibfield  {journal}
  {\bibinfo  {journal} {Phys. Rev. B}\ }\textbf {\bibinfo {volume} {54}},\
  \bibinfo {pages} {10081} (\bibinfo {year} {1996})}\BibitemShut {NoStop}%
\bibitem [{\citenamefont {Chen}\ \emph {et~al.}(2021)\citenamefont {Chen},
  \citenamefont {Wang}, \citenamefont {Swartz}, \citenamefont {Yoon},
  \citenamefont {Hikita}, \citenamefont {Raghu},\ and\ \citenamefont
  {Hwang}}]{chen2021universal}%
  \BibitemOpen
  \bibfield  {author} {\bibinfo {author} {\bibfnamefont {Z.}~\bibnamefont
  {Chen}}, \bibinfo {author} {\bibfnamefont {B.~Y.}\ \bibnamefont {Wang}},
  \bibinfo {author} {\bibfnamefont {A.~G.}\ \bibnamefont {Swartz}}, \bibinfo
  {author} {\bibfnamefont {H.}~\bibnamefont {Yoon}}, \bibinfo {author}
  {\bibfnamefont {Y.}~\bibnamefont {Hikita}}, \bibinfo {author} {\bibfnamefont
  {S.}~\bibnamefont {Raghu}}, \ and\ \bibinfo {author} {\bibfnamefont {H.~Y.}\
  \bibnamefont {Hwang}},\ }\href {\doibase 10.1038/s41535-021-00312-x}
  {\bibfield  {journal} {\bibinfo  {journal} {npj Quantum Materials}\ }\textbf
  {\bibinfo {volume} {6}},\ \bibinfo {pages} {15} (\bibinfo {year}
  {2021})}\BibitemShut {NoStop}%
\bibitem [{\citenamefont {Feigelman}\ \emph {et~al.}(1993)\citenamefont
  {Feigelman}, \citenamefont {Geshkenbein}, \citenamefont {Ioffe},\ and\
  \citenamefont {Larkin}}]{PhysRevB.48.16641}%
  \BibitemOpen
  \bibfield  {author} {\bibinfo {author} {\bibfnamefont {M.~V.}\ \bibnamefont
  {Feigelman}}, \bibinfo {author} {\bibfnamefont {V.~B.}\ \bibnamefont
  {Geshkenbein}}, \bibinfo {author} {\bibfnamefont {L.~B.}\ \bibnamefont
  {Ioffe}}, \ and\ \bibinfo {author} {\bibfnamefont {A.~I.}\ \bibnamefont
  {Larkin}},\ }\href {\doibase 10.1103/PhysRevB.48.16641} {\bibfield  {journal}
  {\bibinfo  {journal} {Phys. Rev. B}\ }\textbf {\bibinfo {volume} {48}},\
  \bibinfo {pages} {16641} (\bibinfo {year} {1993})}\BibitemShut {NoStop}%
\bibitem [{\citenamefont {Motrunich}\ and\ \citenamefont
  {Fisher}(2007)}]{PhysRevB.75.235116}%
  \BibitemOpen
  \bibfield  {author} {\bibinfo {author} {\bibfnamefont {O.~I.}\ \bibnamefont
  {Motrunich}}\ and\ \bibinfo {author} {\bibfnamefont {M.~P.~A.}\ \bibnamefont
  {Fisher}},\ }\href {\doibase 10.1103/PhysRevB.75.235116} {\bibfield
  {journal} {\bibinfo  {journal} {Phys. Rev. B}\ }\textbf {\bibinfo {volume}
  {75}},\ \bibinfo {pages} {235116} (\bibinfo {year} {2007})}\BibitemShut
  {NoStop}%
\bibitem [{\citenamefont {Dalidovich}\ and\ \citenamefont
  {Phillips}(2001)}]{PhysRevB.64.052507}%
  \BibitemOpen
  \bibfield  {author} {\bibinfo {author} {\bibfnamefont {D.}~\bibnamefont
  {Dalidovich}}\ and\ \bibinfo {author} {\bibfnamefont {P.}~\bibnamefont
  {Phillips}},\ }\href {\doibase 10.1103/PhysRevB.64.052507} {\bibfield
  {journal} {\bibinfo  {journal} {Phys. Rev. B}\ }\textbf {\bibinfo {volume}
  {64}},\ \bibinfo {pages} {052507} (\bibinfo {year} {2001})}\BibitemShut
  {NoStop}%
\bibitem [{\citenamefont {Sheng}\ \emph {et~al.}(2008)\citenamefont {Sheng},
  \citenamefont {Motrunich}, \citenamefont {Trebst}, \citenamefont {Gull},\
  and\ \citenamefont {Fisher}}]{PhysRevB.78.054520}%
  \BibitemOpen
  \bibfield  {author} {\bibinfo {author} {\bibfnamefont {D.~N.}\ \bibnamefont
  {Sheng}}, \bibinfo {author} {\bibfnamefont {O.~I.}\ \bibnamefont
  {Motrunich}}, \bibinfo {author} {\bibfnamefont {S.}~\bibnamefont {Trebst}},
  \bibinfo {author} {\bibfnamefont {E.}~\bibnamefont {Gull}}, \ and\ \bibinfo
  {author} {\bibfnamefont {M.~P.~A.}\ \bibnamefont {Fisher}},\ }\href {\doibase
  10.1103/PhysRevB.78.054520} {\bibfield  {journal} {\bibinfo  {journal} {Phys.
  Rev. B}\ }\textbf {\bibinfo {volume} {78}},\ \bibinfo {pages} {054520}
  (\bibinfo {year} {2008})}\BibitemShut {NoStop}%
\bibitem [{\citenamefont {Block}\ \emph {et~al.}(2011)\citenamefont {Block},
  \citenamefont {Mishmash}, \citenamefont {Kaul}, \citenamefont {Sheng},
  \citenamefont {Motrunich},\ and\ \citenamefont
  {Fisher}}]{PhysRevLett.106.046402}%
  \BibitemOpen
  \bibfield  {author} {\bibinfo {author} {\bibfnamefont {M.~S.}\ \bibnamefont
  {Block}}, \bibinfo {author} {\bibfnamefont {R.~V.}\ \bibnamefont {Mishmash}},
  \bibinfo {author} {\bibfnamefont {R.~K.}\ \bibnamefont {Kaul}}, \bibinfo
  {author} {\bibfnamefont {D.~N.}\ \bibnamefont {Sheng}}, \bibinfo {author}
  {\bibfnamefont {O.~I.}\ \bibnamefont {Motrunich}}, \ and\ \bibinfo {author}
  {\bibfnamefont {M.~P.~A.}\ \bibnamefont {Fisher}},\ }\href {\doibase
  10.1103/PhysRevLett.106.046402} {\bibfield  {journal} {\bibinfo  {journal}
  {Phys. Rev. Lett.}\ }\textbf {\bibinfo {volume} {106}},\ \bibinfo {pages}
  {046402} (\bibinfo {year} {2011})}\BibitemShut {NoStop}%
\bibitem [{\citenamefont {Hatsugai}\ and\ \citenamefont
  {Kohmoto}(1992)}]{doi:10.1143/JPSJ.61.2056}%
  \BibitemOpen
  \bibfield  {author} {\bibinfo {author} {\bibfnamefont {Y.}~\bibnamefont
  {Hatsugai}}\ and\ \bibinfo {author} {\bibfnamefont {M.}~\bibnamefont
  {Kohmoto}},\ }\href {\doibase 10.1143/JPSJ.61.2056} {\bibfield  {journal}
  {\bibinfo  {journal} {Journal of the Physical Society of Japan}\ }\textbf
  {\bibinfo {volume} {61}},\ \bibinfo {pages} {2056} (\bibinfo {year}
  {1992})},\ \Eprint
  {http://arxiv.org/abs/https://doi.org/10.1143/JPSJ.61.2056}
  {https://doi.org/10.1143/JPSJ.61.2056} \BibitemShut {NoStop}%
\bibitem [{\citenamefont {BASKARAN}(1991)}]{doi:10.1142/S0217984991000782}%
  \BibitemOpen
  \bibfield  {author} {\bibinfo {author} {\bibfnamefont {G.}~\bibnamefont
  {BASKARAN}},\ }\href {\doibase 10.1142/S0217984991000782} {\bibfield
  {journal} {\bibinfo  {journal} {Modern Physics Letters B}\ }\textbf {\bibinfo
  {volume} {05}},\ \bibinfo {pages} {643} (\bibinfo {year} {1991})},\ \Eprint
  {http://arxiv.org/abs/https://doi.org/10.1142/S0217984991000782}
  {https://doi.org/10.1142/S0217984991000782} \BibitemShut {NoStop}%
\bibitem [{\citenamefont {Sachdev}\ and\ \citenamefont
  {Ye}(1993)}]{PhysRevLett.70.3339}%
  \BibitemOpen
  \bibfield  {author} {\bibinfo {author} {\bibfnamefont {S.}~\bibnamefont
  {Sachdev}}\ and\ \bibinfo {author} {\bibfnamefont {J.}~\bibnamefont {Ye}},\
  }\href {\doibase 10.1103/PhysRevLett.70.3339} {\bibfield  {journal} {\bibinfo
   {journal} {Phys. Rev. Lett.}\ }\textbf {\bibinfo {volume} {70}},\ \bibinfo
  {pages} {3339} (\bibinfo {year} {1993})}\BibitemShut {NoStop}%
\bibitem [{\citenamefont {Maldacena}\ and\ \citenamefont
  {Stanford}(2016)}]{PhysRevD.94.106002}%
  \BibitemOpen
  \bibfield  {author} {\bibinfo {author} {\bibfnamefont {J.}~\bibnamefont
  {Maldacena}}\ and\ \bibinfo {author} {\bibfnamefont {D.}~\bibnamefont
  {Stanford}},\ }\href {\doibase 10.1103/PhysRevD.94.106002} {\bibfield
  {journal} {\bibinfo  {journal} {Phys. Rev. D}\ }\textbf {\bibinfo {volume}
  {94}},\ \bibinfo {pages} {106002} (\bibinfo {year} {2016})}\BibitemShut
  {NoStop}%
\bibitem [{\citenamefont {Chowdhury}\ \emph {et~al.}(2022)\citenamefont
  {Chowdhury}, \citenamefont {Georges}, \citenamefont {Parcollet},\ and\
  \citenamefont {Sachdev}}]{RevModPhys.94.035004}%
  \BibitemOpen
  \bibfield  {author} {\bibinfo {author} {\bibfnamefont {D.}~\bibnamefont
  {Chowdhury}}, \bibinfo {author} {\bibfnamefont {A.}~\bibnamefont {Georges}},
  \bibinfo {author} {\bibfnamefont {O.}~\bibnamefont {Parcollet}}, \ and\
  \bibinfo {author} {\bibfnamefont {S.}~\bibnamefont {Sachdev}},\ }\href
  {\doibase 10.1103/RevModPhys.94.035004} {\bibfield  {journal} {\bibinfo
  {journal} {Rev. Mod. Phys.}\ }\textbf {\bibinfo {volume} {94}},\ \bibinfo
  {pages} {035004} (\bibinfo {year} {2022})}\BibitemShut {NoStop}%
\bibitem [{\citenamefont {Zhou}\ \emph {et~al.}(2017)\citenamefont {Zhou},
  \citenamefont {Kanoda},\ and\ \citenamefont {Ng}}]{RevModPhys.89.025003}%
  \BibitemOpen
  \bibfield  {author} {\bibinfo {author} {\bibfnamefont {Y.}~\bibnamefont
  {Zhou}}, \bibinfo {author} {\bibfnamefont {K.}~\bibnamefont {Kanoda}}, \ and\
  \bibinfo {author} {\bibfnamefont {T.-K.}\ \bibnamefont {Ng}},\ }\href
  {\doibase 10.1103/RevModPhys.89.025003} {\bibfield  {journal} {\bibinfo
  {journal} {Rev. Mod. Phys.}\ }\textbf {\bibinfo {volume} {89}},\ \bibinfo
  {pages} {025003} (\bibinfo {year} {2017})}\BibitemShut {NoStop}%
\bibitem [{\citenamefont {Prosko}\ \emph {et~al.}(2017)\citenamefont {Prosko},
  \citenamefont {Lee},\ and\ \citenamefont {Maciejko}}]{PhysRevB.96.205104}%
  \BibitemOpen
  \bibfield  {author} {\bibinfo {author} {\bibfnamefont {C.}~\bibnamefont
  {Prosko}}, \bibinfo {author} {\bibfnamefont {S.-P.}\ \bibnamefont {Lee}}, \
  and\ \bibinfo {author} {\bibfnamefont {J.}~\bibnamefont {Maciejko}},\ }\href
  {\doibase 10.1103/PhysRevB.96.205104} {\bibfield  {journal} {\bibinfo
  {journal} {Phys. Rev. B}\ }\textbf {\bibinfo {volume} {96}},\ \bibinfo
  {pages} {205104} (\bibinfo {year} {2017})}\BibitemShut {NoStop}%
\bibitem [{\citenamefont {Zhong}\ \emph {et~al.}(2013)\citenamefont {Zhong},
  \citenamefont {Wang},\ and\ \citenamefont {Luo}}]{PhysRevB.88.045109}%
  \BibitemOpen
  \bibfield  {author} {\bibinfo {author} {\bibfnamefont {Y.}~\bibnamefont
  {Zhong}}, \bibinfo {author} {\bibfnamefont {Y.-F.}\ \bibnamefont {Wang}}, \
  and\ \bibinfo {author} {\bibfnamefont {H.-G.}\ \bibnamefont {Luo}},\ }\href
  {\doibase 10.1103/PhysRevB.88.045109} {\bibfield  {journal} {\bibinfo
  {journal} {Phys. Rev. B}\ }\textbf {\bibinfo {volume} {88}},\ \bibinfo
  {pages} {045109} (\bibinfo {year} {2013})}\BibitemShut {NoStop}%
\bibitem [{\citenamefont {Smith}\ \emph {et~al.}(2017)\citenamefont {Smith},
  \citenamefont {Knolle}, \citenamefont {Kovrizhin},\ and\ \citenamefont
  {Moessner}}]{PhysRevLett.118.266601}%
  \BibitemOpen
  \bibfield  {author} {\bibinfo {author} {\bibfnamefont {A.}~\bibnamefont
  {Smith}}, \bibinfo {author} {\bibfnamefont {J.}~\bibnamefont {Knolle}},
  \bibinfo {author} {\bibfnamefont {D.~L.}\ \bibnamefont {Kovrizhin}}, \ and\
  \bibinfo {author} {\bibfnamefont {R.}~\bibnamefont {Moessner}},\ }\href
  {\doibase 10.1103/PhysRevLett.118.266601} {\bibfield  {journal} {\bibinfo
  {journal} {Phys. Rev. Lett.}\ }\textbf {\bibinfo {volume} {118}},\ \bibinfo
  {pages} {266601} (\bibinfo {year} {2017})}\BibitemShut {NoStop}%
\bibitem [{\citenamefont {Chen}\ \emph
  {et~al.}(2018{\natexlab{b}})\citenamefont {Chen}, \citenamefont {Li},\ and\
  \citenamefont {Ng}}]{PhysRevLett.120.046401}%
  \BibitemOpen
  \bibfield  {author} {\bibinfo {author} {\bibfnamefont {Z.}~\bibnamefont
  {Chen}}, \bibinfo {author} {\bibfnamefont {X.}~\bibnamefont {Li}}, \ and\
  \bibinfo {author} {\bibfnamefont {T.~K.}\ \bibnamefont {Ng}},\ }\href
  {\doibase 10.1103/PhysRevLett.120.046401} {\bibfield  {journal} {\bibinfo
  {journal} {Phys. Rev. Lett.}\ }\textbf {\bibinfo {volume} {120}},\ \bibinfo
  {pages} {046401} (\bibinfo {year} {2018}{\natexlab{b}})}\BibitemShut
  {NoStop}%
\bibitem [{\citenamefont {Kitaev}(2003)}]{KITAEV20032}%
  \BibitemOpen
  \bibfield  {author} {\bibinfo {author} {\bibfnamefont {A.}~\bibnamefont
  {Kitaev}},\ }\href {\doibase https://doi.org/10.1016/S0003-4916(02)00018-0}
  {\bibfield  {journal} {\bibinfo  {journal} {Annals of Physics}\ }\textbf
  {\bibinfo {volume} {303}},\ \bibinfo {pages} {2} (\bibinfo {year}
  {2003})}\BibitemShut {NoStop}%
\bibitem [{\citenamefont {Kitaev}(2006)}]{KITAEV20062}%
  \BibitemOpen
  \bibfield  {author} {\bibinfo {author} {\bibfnamefont {A.}~\bibnamefont
  {Kitaev}},\ }\href {\doibase https://doi.org/10.1016/j.aop.2005.10.005}
  {\bibfield  {journal} {\bibinfo  {journal} {Annals of Physics}\ }\textbf
  {\bibinfo {volume} {321}},\ \bibinfo {pages} {2} (\bibinfo {year} {2006})},\
  \bibinfo {note} {january Special Issue}\BibitemShut {NoStop}%
\bibitem [{\citenamefont {Continentino}\ and\ \citenamefont
  {Coutinho-Filho}(1994)}]{CONTINENTINO1994619}%
  \BibitemOpen
  \bibfield  {author} {\bibinfo {author} {\bibfnamefont {M.~A.}\ \bibnamefont
  {Continentino}}\ and\ \bibinfo {author} {\bibfnamefont {M.~D.}\ \bibnamefont
  {Coutinho-Filho}},\ }\href {\doibase
  https://doi.org/10.1016/0038-1098(94)90533-9} {\bibfield  {journal} {\bibinfo
   {journal} {Solid State Communications}\ }\textbf {\bibinfo {volume} {90}},\
  \bibinfo {pages} {619} (\bibinfo {year} {1994})}\BibitemShut {NoStop}%
\bibitem [{\citenamefont {Hatsugai}\ \emph {et~al.}(1996)\citenamefont
  {Hatsugai}, \citenamefont {Kohmoto}, \citenamefont {Koma},\ and\
  \citenamefont {Wu}}]{PhysRevB.54.5358}%
  \BibitemOpen
  \bibfield  {author} {\bibinfo {author} {\bibfnamefont {Y.}~\bibnamefont
  {Hatsugai}}, \bibinfo {author} {\bibfnamefont {M.}~\bibnamefont {Kohmoto}},
  \bibinfo {author} {\bibfnamefont {T.}~\bibnamefont {Koma}}, \ and\ \bibinfo
  {author} {\bibfnamefont {Y.-S.}\ \bibnamefont {Wu}},\ }\href {\doibase
  10.1103/PhysRevB.54.5358} {\bibfield  {journal} {\bibinfo  {journal} {Phys.
  Rev. B}\ }\textbf {\bibinfo {volume} {54}},\ \bibinfo {pages} {5358}
  (\bibinfo {year} {1996})}\BibitemShut {NoStop}%
\bibitem [{\citenamefont {Phillips}\ \emph {et~al.}(2018)\citenamefont
  {Phillips}, \citenamefont {Setty},\ and\ \citenamefont
  {Zhang}}]{PhysRevB.97.195102}%
  \BibitemOpen
  \bibfield  {author} {\bibinfo {author} {\bibfnamefont {P.~W.}\ \bibnamefont
  {Phillips}}, \bibinfo {author} {\bibfnamefont {C.}~\bibnamefont {Setty}}, \
  and\ \bibinfo {author} {\bibfnamefont {S.}~\bibnamefont {Zhang}},\ }\href
  {\doibase 10.1103/PhysRevB.97.195102} {\bibfield  {journal} {\bibinfo
  {journal} {Phys. Rev. B}\ }\textbf {\bibinfo {volume} {97}},\ \bibinfo
  {pages} {195102} (\bibinfo {year} {2018})}\BibitemShut {NoStop}%
\bibitem [{\citenamefont {Yeo}\ and\ \citenamefont
  {Phillips}(2019)}]{PhysRevD.99.094030}%
  \BibitemOpen
  \bibfield  {author} {\bibinfo {author} {\bibfnamefont {L.}~\bibnamefont
  {Yeo}}\ and\ \bibinfo {author} {\bibfnamefont {P.~W.}\ \bibnamefont
  {Phillips}},\ }\href {\doibase 10.1103/PhysRevD.99.094030} {\bibfield
  {journal} {\bibinfo  {journal} {Phys. Rev. D}\ }\textbf {\bibinfo {volume}
  {99}},\ \bibinfo {pages} {094030} (\bibinfo {year} {2019})}\BibitemShut
  {NoStop}%
\bibitem [{\citenamefont {Zhu}\ \emph {et~al.}(2021)\citenamefont {Zhu},
  \citenamefont {Li}, \citenamefont {Han},\ and\ \citenamefont
  {Wang}}]{PhysRevB.103.024514}%
  \BibitemOpen
  \bibfield  {author} {\bibinfo {author} {\bibfnamefont {H.-S.}\ \bibnamefont
  {Zhu}}, \bibinfo {author} {\bibfnamefont {Z.}~\bibnamefont {Li}}, \bibinfo
  {author} {\bibfnamefont {Q.}~\bibnamefont {Han}}, \ and\ \bibinfo {author}
  {\bibfnamefont {Z.~D.}\ \bibnamefont {Wang}},\ }\href {\doibase
  10.1103/PhysRevB.103.024514} {\bibfield  {journal} {\bibinfo  {journal}
  {Phys. Rev. B}\ }\textbf {\bibinfo {volume} {103}},\ \bibinfo {pages}
  {024514} (\bibinfo {year} {2021})}\BibitemShut {NoStop}%
\bibitem [{\citenamefont {Yang}(2021)}]{PhysRevB.103.024529}%
  \BibitemOpen
  \bibfield  {author} {\bibinfo {author} {\bibfnamefont {K.}~\bibnamefont
  {Yang}},\ }\href {\doibase 10.1103/PhysRevB.103.024529} {\bibfield  {journal}
  {\bibinfo  {journal} {Phys. Rev. B}\ }\textbf {\bibinfo {volume} {103}},\
  \bibinfo {pages} {024529} (\bibinfo {year} {2021})}\BibitemShut {NoStop}%
\bibitem [{\citenamefont {Zhao}\ \emph {et~al.}(2022)\citenamefont {Zhao},
  \citenamefont {Yeo}, \citenamefont {Huang},\ and\ \citenamefont
  {Phillips}}]{PhysRevB.105.184509}%
  \BibitemOpen
  \bibfield  {author} {\bibinfo {author} {\bibfnamefont {J.}~\bibnamefont
  {Zhao}}, \bibinfo {author} {\bibfnamefont {L.}~\bibnamefont {Yeo}}, \bibinfo
  {author} {\bibfnamefont {E.~W.}\ \bibnamefont {Huang}}, \ and\ \bibinfo
  {author} {\bibfnamefont {P.~W.}\ \bibnamefont {Phillips}},\ }\href {\doibase
  10.1103/PhysRevB.105.184509} {\bibfield  {journal} {\bibinfo  {journal}
  {Phys. Rev. B}\ }\textbf {\bibinfo {volume} {105}},\ \bibinfo {pages}
  {184509} (\bibinfo {year} {2022})}\BibitemShut {NoStop}%
\bibitem [{\citenamefont {Huang}\ \emph {et~al.}(2022)\citenamefont {Huang},
  \citenamefont {Nave},\ and\ \citenamefont {Phillips}}]{Huang2022}%
  \BibitemOpen
  \bibfield  {author} {\bibinfo {author} {\bibfnamefont {E.~W.}\ \bibnamefont
  {Huang}}, \bibinfo {author} {\bibfnamefont {G.~L.}\ \bibnamefont {Nave}}, \
  and\ \bibinfo {author} {\bibfnamefont {P.~W.}\ \bibnamefont {Phillips}},\
  }\href {\doibase 10.1038/s41567-022-01529-8} {\bibfield  {journal} {\bibinfo
  {journal} {Nature Physics}\ }\textbf {\bibinfo {volume} {18}},\ \bibinfo
  {pages} {511} (\bibinfo {year} {2022})}\BibitemShut {NoStop}%
\bibitem [{\citenamefont {Mai}\ \emph {et~al.}(2023)\citenamefont {Mai},
  \citenamefont {Feldman},\ and\ \citenamefont
  {Phillips}}]{PhysRevResearch.5.013162}%
  \BibitemOpen
  \bibfield  {author} {\bibinfo {author} {\bibfnamefont {P.}~\bibnamefont
  {Mai}}, \bibinfo {author} {\bibfnamefont {B.~E.}\ \bibnamefont {Feldman}}, \
  and\ \bibinfo {author} {\bibfnamefont {P.~W.}\ \bibnamefont {Phillips}},\
  }\href {\doibase 10.1103/PhysRevResearch.5.013162} {\bibfield  {journal}
  {\bibinfo  {journal} {Phys. Rev. Res.}\ }\textbf {\bibinfo {volume} {5}},\
  \bibinfo {pages} {013162} (\bibinfo {year} {2023})}\BibitemShut {NoStop}%
\bibitem [{\citenamefont {Li}\ \emph {et~al.}(2022)\citenamefont {Li},
  \citenamefont {Mishra}, \citenamefont {Zhou},\ and\ \citenamefont
  {Zhang}}]{Li_2022}%
  \BibitemOpen
  \bibfield  {author} {\bibinfo {author} {\bibfnamefont {Y.}~\bibnamefont
  {Li}}, \bibinfo {author} {\bibfnamefont {V.}~\bibnamefont {Mishra}}, \bibinfo
  {author} {\bibfnamefont {Y.}~\bibnamefont {Zhou}}, \ and\ \bibinfo {author}
  {\bibfnamefont {F.-C.}\ \bibnamefont {Zhang}},\ }\href {\doibase
  10.1088/1367-2630/ac9548} {\bibfield  {journal} {\bibinfo  {journal} {New
  Journal of Physics}\ }\textbf {\bibinfo {volume} {24}},\ \bibinfo {pages}
  {103019} (\bibinfo {year} {2022})}\BibitemShut {NoStop}%
\bibitem [{\citenamefont {Setty}(2020)}]{PhysRevB.101.184506}%
  \BibitemOpen
  \bibfield  {author} {\bibinfo {author} {\bibfnamefont {C.}~\bibnamefont
  {Setty}},\ }\href {\doibase 10.1103/PhysRevB.101.184506} {\bibfield
  {journal} {\bibinfo  {journal} {Phys. Rev. B}\ }\textbf {\bibinfo {volume}
  {101}},\ \bibinfo {pages} {184506} (\bibinfo {year} {2020})}\BibitemShut
  {NoStop}%
\bibitem [{\citenamefont {Setty}(2021{\natexlab{a}})}]{PhysRevB.103.014501}%
  \BibitemOpen
  \bibfield  {author} {\bibinfo {author} {\bibfnamefont {C.}~\bibnamefont
  {Setty}},\ }\href {\doibase 10.1103/PhysRevB.103.014501} {\bibfield
  {journal} {\bibinfo  {journal} {Phys. Rev. B}\ }\textbf {\bibinfo {volume}
  {103}},\ \bibinfo {pages} {014501} (\bibinfo {year}
  {2021}{\natexlab{a}})}\BibitemShut {NoStop}%
\bibitem [{\citenamefont {Phillips}\ \emph {et~al.}(2020)\citenamefont
  {Phillips}, \citenamefont {Yeo},\ and\ \citenamefont
  {Huang}}]{Phillips_2020}%
  \BibitemOpen
  \bibfield  {author} {\bibinfo {author} {\bibfnamefont {P.~W.}\ \bibnamefont
  {Phillips}}, \bibinfo {author} {\bibfnamefont {L.}~\bibnamefont {Yeo}}, \
  and\ \bibinfo {author} {\bibfnamefont {E.~W.}\ \bibnamefont {Huang}},\ }\href
  {\doibase 10.1038/s41567-020-0988-4} {\bibfield  {journal} {\bibinfo
  {journal} {Nature Physics}\ } (\bibinfo {year} {2020}),\
  10.1038/s41567-020-0988-4}\BibitemShut {NoStop}%
\bibitem [{\citenamefont {Setty}(2021{\natexlab{b}})}]{arXiv:2105.15205}%
  \BibitemOpen
  \bibfield  {author} {\bibinfo {author} {\bibfnamefont {C.}~\bibnamefont
  {Setty}},\ }\href {\doibase 10.48550/arXiv.2105.15205} {\  (\bibinfo {year}
  {2021}{\natexlab{b}}),\ 10.48550/arXiv.2105.15205}\BibitemShut {NoStop}%
\bibitem [{\citenamefont {Zhong}(2022)}]{PhysRevB.106.155119}%
  \BibitemOpen
  \bibfield  {author} {\bibinfo {author} {\bibfnamefont {Y.}~\bibnamefont
  {Zhong}},\ }\href {\doibase 10.1103/PhysRevB.106.155119} {\bibfield
  {journal} {\bibinfo  {journal} {Phys. Rev. B}\ }\textbf {\bibinfo {volume}
  {106}},\ \bibinfo {pages} {155119} (\bibinfo {year} {2022})}\BibitemShut
  {NoStop}%
\bibitem [{\citenamefont {Zhao}\ \emph {et~al.}(2023)\citenamefont {Zhao},
  \citenamefont {Yang}, \citenamefont {Luo},\ and\ \citenamefont
  {Zhong}}]{arXiv:2303.00926}%
  \BibitemOpen
  \bibfield  {author} {\bibinfo {author} {\bibfnamefont {M.}~\bibnamefont
  {Zhao}}, \bibinfo {author} {\bibfnamefont {W.-W.}\ \bibnamefont {Yang}},
  \bibinfo {author} {\bibfnamefont {H.-G.}\ \bibnamefont {Luo}}, \ and\
  \bibinfo {author} {\bibfnamefont {Y.}~\bibnamefont {Zhong}},\ }\href
  {\doibase 10.48550/arXiv.2303.00926} {\  (\bibinfo {year} {2023}),\
  10.48550/arXiv.2303.00926}\BibitemShut {NoStop}%
\bibitem [{\citenamefont {Zeng}\ \emph {et~al.}(2019)\citenamefont {Zeng},
  \citenamefont {Chen}, \citenamefont {Zhou},\ and\ \citenamefont
  {Wen}}]{Wen2019}%
  \BibitemOpen
  \bibfield  {author} {\bibinfo {author} {\bibfnamefont {B.}~\bibnamefont
  {Zeng}}, \bibinfo {author} {\bibfnamefont {X.}~\bibnamefont {Chen}}, \bibinfo
  {author} {\bibfnamefont {D.-L.}\ \bibnamefont {Zhou}}, \ and\ \bibinfo
  {author} {\bibfnamefont {X.-G.}\ \bibnamefont {Wen}},\ }\href {\doibase
  10.1007/978-1-4939-9084-9} {\emph {\bibinfo {title} {Quantum Information
  Meets Quantum Matter}}}\ (\bibinfo  {publisher} {Springer New York},\
  \bibinfo {year} {2019})\BibitemShut {NoStop}%
\bibitem [{\citenamefont {Scalapino}\ \emph {et~al.}(1992)\citenamefont
  {Scalapino}, \citenamefont {White},\ and\ \citenamefont
  {Zhang}}]{PhysRevLett.68.2830}%
  \BibitemOpen
  \bibfield  {author} {\bibinfo {author} {\bibfnamefont {D.~J.}\ \bibnamefont
  {Scalapino}}, \bibinfo {author} {\bibfnamefont {S.~R.}\ \bibnamefont
  {White}}, \ and\ \bibinfo {author} {\bibfnamefont {S.~C.}\ \bibnamefont
  {Zhang}},\ }\href {\doibase 10.1103/PhysRevLett.68.2830} {\bibfield
  {journal} {\bibinfo  {journal} {Phys. Rev. Lett.}\ }\textbf {\bibinfo
  {volume} {68}},\ \bibinfo {pages} {2830} (\bibinfo {year}
  {1992})}\BibitemShut {NoStop}%
\bibitem [{\citenamefont {Continentino}(2017)}]{Mucio}%
  \BibitemOpen
  \bibfield  {author} {\bibinfo {author} {\bibfnamefont {M.~A.}\ \bibnamefont
  {Continentino}},\ }\href {\doibase 10.1017/CBO9781316576854} {\emph {\bibinfo
  {title} {Quantum Scaling in Many-Body Systems: An Approach to Quantum Phase
  Transitions}}}\ (\bibinfo  {publisher} {Cambridge University Press},\
  \bibinfo {year} {2017})\BibitemShut {NoStop}%
\bibitem [{\citenamefont {Vitoriano}\ \emph {et~al.}(2000)\citenamefont
  {Vitoriano}, \citenamefont {Bejan}, \citenamefont {Mac\^edo},\ and\
  \citenamefont {Coutinho-Filho}}]{PhysRevB.61.7941}%
  \BibitemOpen
  \bibfield  {author} {\bibinfo {author} {\bibfnamefont {C.}~\bibnamefont
  {Vitoriano}}, \bibinfo {author} {\bibfnamefont {L.~B.}\ \bibnamefont
  {Bejan}}, \bibinfo {author} {\bibfnamefont {A.~M.~S.}\ \bibnamefont
  {Mac\^edo}}, \ and\ \bibinfo {author} {\bibfnamefont {M.~D.}\ \bibnamefont
  {Coutinho-Filho}},\ }\href {\doibase 10.1103/PhysRevB.61.7941} {\bibfield
  {journal} {\bibinfo  {journal} {Phys. Rev. B}\ }\textbf {\bibinfo {volume}
  {61}},\ \bibinfo {pages} {7941} (\bibinfo {year} {2000})}\BibitemShut
  {NoStop}%
\bibitem [{\citenamefont {Coleman}(2015)}]{Coleman}%
  \BibitemOpen
  \bibfield  {author} {\bibinfo {author} {\bibfnamefont {P.}~\bibnamefont
  {Coleman}},\ }\href {\doibase 10.1017/CBO9781139020916} {\emph {\bibinfo
  {title} {Introduction to many-body physics}}}\ (\bibinfo  {publisher}
  {Cambridge University Press},\ \bibinfo {year} {2015})\BibitemShut {NoStop}%
\bibitem [{\citenamefont {Alet}\ and\ \citenamefont
  {S\o{}rensen}(2004)}]{PhysRevB.70.024513}%
  \BibitemOpen
  \bibfield  {author} {\bibinfo {author} {\bibfnamefont {F.}~\bibnamefont
  {Alet}}\ and\ \bibinfo {author} {\bibfnamefont {E.~S.}\ \bibnamefont
  {S\o{}rensen}},\ }\href {\doibase 10.1103/PhysRevB.70.024513} {\bibfield
  {journal} {\bibinfo  {journal} {Phys. Rev. B}\ }\textbf {\bibinfo {volume}
  {70}},\ \bibinfo {pages} {024513} (\bibinfo {year} {2004})}\BibitemShut
  {NoStop}%
\bibitem [{\citenamefont {K\"uhner}\ \emph {et~al.}(2000)\citenamefont
  {K\"uhner}, \citenamefont {White},\ and\ \citenamefont
  {Monien}}]{PhysRevB.61.12474}%
  \BibitemOpen
  \bibfield  {author} {\bibinfo {author} {\bibfnamefont {T.~D.}\ \bibnamefont
  {K\"uhner}}, \bibinfo {author} {\bibfnamefont {S.~R.}\ \bibnamefont {White}},
  \ and\ \bibinfo {author} {\bibfnamefont {H.}~\bibnamefont {Monien}},\ }\href
  {\doibase 10.1103/PhysRevB.61.12474} {\bibfield  {journal} {\bibinfo
  {journal} {Phys. Rev. B}\ }\textbf {\bibinfo {volume} {61}},\ \bibinfo
  {pages} {12474} (\bibinfo {year} {2000})}\BibitemShut {NoStop}%
\end{thebibliography}%

\end{document}